\documentclass[10pt,conference]{IEEEtran}
\IEEEoverridecommandlockouts
\usepackage{cite}
\usepackage{amsmath,amsthm,amssymb,amsfonts,braket}
\usepackage{algorithmicx}
\usepackage{graphicx}
\usepackage{textcomp}
\usepackage{xcolor}
\usepackage{algorithm,balance}
\usepackage[noend]{algpseudocode}
\usepackage{breqn}
\usepackage{physics}
\usepackage{subfigure}
\usepackage{xcolor}
\usepackage{mathtools}
\usepackage{lipsum}

\newtheorem {theorem} {Theorem}

\newcommand\note[3]{{\textcolor{#1}{[\textsf{#2}: #3]}}}

\newcommand{\bing}[1]{\note{blue}{BW}{#1}}

\newcommand{\trd}[1]{\left|\left|#1\right|\right|}

\algnewcommand\algorithmicforeach{\textbf{for each}}
\algdef{S}[FOR]{ForEach}[1]{\algorithmicforeach\ #1\ \algorithmicdo}

\def\BibTeX{{\rm B\kern-.05em{\sc i\kern-.025em b}\kern-.08em
    T\kern-.1667em\lower.7ex\hbox{E}\kern-.125emX}}

\graphicspath{ {./images/} }

\makeatletter
\newcommand{\linebreakand}{%
  \end{@IEEEauthorhalign}
  \hfill\mbox{}\par
  \mbox{}\hfill\begin{@IEEEauthorhalign}
}
\makeatother

\usepackage{epsfig}
\usepackage{algorithm}

\newcommand{\remove}[1]{}

\newtheorem{lemma}{Lemma}
\newtheorem{corollary}{Corollary}

\newcommand{\ls}[1]
   {\dimen0=\fontdimen6\the\font
    \lineskip=#1\dimen0
    \advance\lineskip.5\fontdimen5\the\font
    \advance\lineskip-\dimen0
    \lineskiplimit=.9\lineskip
    \baselineskip=\lineskip
    \advance\baselineskip\dimen0
    \normallineskip\lineskip
    \normallineskiplimit\lineskiplimit
    \normalbaselineskip\baselineskip
    \ignorespaces
   }

\newcommand{\singlefig}[3]{
\begin{figure}
\centerline{
    \setlength{\epsfysize}{0.25\textwidth}
 \epsffile{\Figdir#1}
} \caption{#2} \label{fig:#3}
\end{figure}
}

\newcommand{\Figdir}{./}

\begin{document}


\title{
Blockwise Key Distillation in Satellite-based Quantum Key Distribution
}

\author{
\IEEEauthorblockN{Minu J. Bae}
\IEEEauthorblockA{
\textit{University of Connecticut}\\
Storrs CT, USA \\
minwoo.bae@uconn.edu}
\and
\IEEEauthorblockN{Nitish K. Panigrahy}
\IEEEauthorblockA{
\textit{University of Massachusetts}\\
Amherst MA, USA \\
nitish@cs.umass.edu \\
}
\and
\IEEEauthorblockN{Prajit Dhara}
\IEEEauthorblockA{
\textit{University of Arizona}\\ 
Tucson AZ, USA \\
prajitd@email.arizona.edu \\
}\and
\IEEEauthorblockN{Walter O. Krawec}
\IEEEauthorblockA{
\textit{University of Connecticut}\\
Storrs CT, USA\\
walter.krawec@uconn.edu
}
\linebreakand 
\IEEEauthorblockN{Alexander Russell}
\IEEEauthorblockA{
\textit{University of Connecticut}\\
Storrs CT, USA\\
acr@uconn.edu
}
\and
\IEEEauthorblockN{Don Towsley}
\IEEEauthorblockA{
\textit{University of Massachusetts}\\
Amherst MA, USA \\
towsley@cs.umass.edu
}
\and
\IEEEauthorblockN{Bing Wang}
\IEEEauthorblockA{
\textit{University of Connecticut}\\
Storrs CT, USA \\
bing@uconn.edu}
}


\maketitle

\begin{abstract} 
Free-space satellite communication has significantly lower photon loss than  terrestrial communication via optical fibers. Satellite-based quantum key distribution (QKD) leverages this advantage and provides a promising direction in achieving long-distance  inter-continental QKD. Satellite channels, however, can be highly dynamic, due to various environmental factors and time-of-the-day effects, leading to heterogeneous noises over time. In this paper, we compare two key distillation techniques for satellite-based QKD. One is the traditional {\em non-blockwise} strategy that treats all the signals as a whole; the other is a {\em blockwise} strategy that divides the signals into individual blocks that have similar noise characteristics and processes them independently. Through extensive simulation in a wide range of settings, we show trend in optimal parameter choices and when one strategy provides better key generation rates than the other. Our results show that the blockwise strategy can lead to up to $5\%$ key rate improvement (leading to on average $1.9\times10^{7}$ more key bits per day) when considering two types of blocks, i.e., for nighttime and daytime, respectively. The blockwise strategy only requires changes in the classical post-processing stage of QKD and can be easily deployed in existing satellite systems.

\end{abstract}

\section{Introduction}


Quantum cryptography, and specifically Quantum Key Distribution (QKD), holds several promising benefits. In particular, it has the ability to achieve certain cryptographic tasks without relying on computational assumptions, unlike much of our current-day secure communication infrastructure based on public key systems~\cite{Scarani09:QKD-survey,Pirandola19:QKD-Survey,Diamanti16:Practical}. However, several challenges remain, limiting its effectiveness and the rate of adoption of this technology.  One challenge is that long distance quantum communication links are required, while greater key generation rates are necessary to allow for either faster refreshing of AES keys or, hopefully, the ability to stream a true one-time-pad at a rate fast enough to keep up with the communication stream.  Due to the exponential loss in fiber channels~\cite{Svelto10:Principles,Kaushal17:optical}, many researchers are turning to the study of satellite-based quantum communication to solve this long-distance quantum communication problem, leveraging free-space satellite communication that has much lower loss than fiber channels.
A satellite can help to distribute long-range entangled pairs to two ground stations, thus building a QKD network over much longer distances than a point-to-point ground fiber network (without repeaters) could achieve on its own \cite{bedington2017progress}.

Several experimental demonstrations of quantum communication through satellites have been conducted recently \cite{ursin2007entanglement
, yin2012quantum
, ma2012quantum
, wang2013direct
, nauerth2013air
, yin2013experimental
, vallone2015experimental, liao2017satellite, liao2018satellite, hosseinidehaj2018satellite, sidhu2021advances, pirandola2021satellite,mastriani2021satellite, sidhu2022finite}, showing their technological feasibility.  Despite this interest, several questions remain, especially in terms of optimizing overall QKD system performance and speed.  Since the hardware of these systems would be difficult to change after launch, it is important to investigate what can be done on the \emph{classical} stage of the protocol, without forcing users to invest or install new quantum hardware.  Every QKD protocol consists of two stages: a quantum communication stage and a classical post-processing stage.  The first is the only one that requires quantum-capable hardware; the second involves only classical communication and can more easily be altered than the first.

In this work, we investigate the classical post-processing stage of a standard QKD protocol, specifically BB84~\cite{BB14:QKD} (or, rather, the entanglement based version E91~\cite{ekert1991quantum}) in an attempt to maximize the performance of a satellite system, without altering the quantum layer of the network.  Satellite channels show dynamic environmental circumstances due to various time and weather conditions \cite{elser2015satellite, vasylyev2017free, liao2017long, liorni2019satellite, pirandola2021satellite}. For instance, nighttime and daytime have different background/thermal photons that impact the fidelity of entangled photons or their loss, while weather conditions in the atmospheric layers increase noise on entangled photons. So, we must carefully consider the impact of these factors on QKD. Next, it is necessary to determine the optimal pump power to generate the right entangled photon pairs and a sampling rate to estimate noise in the finite key analysis.

In this work, we compare two classical post-processing strategies for satellite systems.  In one instance, which we call ``blockwise post-processing,'' we divide an entire signal into several individual ``blocks'' and process them  independently; these blocks should have similar noise/loss characteristics.  
The other strategy is the more traditional method of treating the entire quantum signal as a single unit and processing accordingly (which we call ``non-blockwise post-processing'').  
For both post-processing methods, we investigate optimal parameter settings for various satellite configurations and operating conditions.

We make several contributions in this work.  To our knowledge, we are the first to evaluate and compare
these two different post-processing methods in both the finite key and asymptotic scenario and under various satellite operating conditions.
%
We also conduct a rigorous evaluation of QKD satellite operation using extensive simulations with realistic noise and loss models, 
showing trends in optimal parameter choices and when, exactly, the two different post-processing methods should be used to optimize overall key generation rates. For instance, we show that the blockwise scheme can lead to 5\% higher key rate than the non-blockwise scheme when the satellite is at a high altitude, leading to on average $1.9\times10^{7}$ more key bits per day.

We comment that all our investigations are  on the classical stage of the QKD protocol, and any alterations which our work suggests that may be beneficial to a satellite QKD system, can be easily adopted by the current systems, and easily added after a satellite's launch. In addition, while focusing on satellite-based QKD, our findings also apply to terrestrial QKD network scenarios where the raw key bits have significant dynamics, e.g., because they are created over disparate network paths.

\section{Preliminaries} \label{sec:prelim}

\begin{figure}[t]
    \centering  \includegraphics[width=0.48\textwidth]{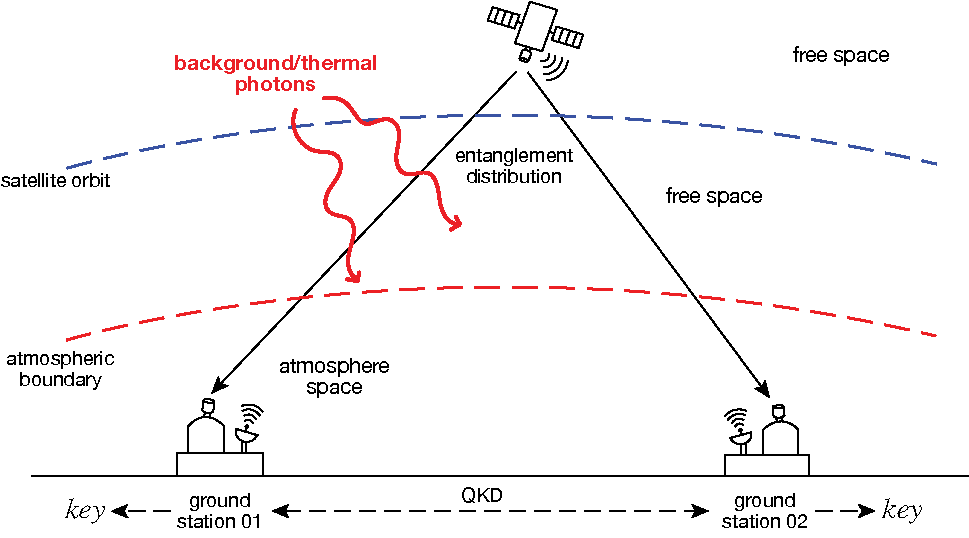}
    \caption{Dual downlink based QKD from one satellite to two ground stations. }
    \label{fig:sat-QKD}
\end{figure}

\subsection{Satellite-based QKD} \label{sec:sate-QKD}

We consider a satellite that orbits around the Earth at a certain altitude. The satellite has photon sources that generate entangled pairs, and sends them to a pair of ground stations. Specifically, for each entangled pair, the satellite transmits one photon in the pair to one ground station using a down-link optical channel, thus creating a dual downlink entanglement distribution as shown in Fig.~\ref{fig:sat-QKD}. The two ground stations run an entanglement based protocol (e.g., E91~\cite{ekert1991quantum}) for QKD. 

In the rest of this paper, we assume that the two ground stations are located on the equator. The satellite orbits the Earth in a west-to-east direction above the equator, in alignment with the Earth's rotation. We focus on low-earth-orbit (LEO) satellites (i.e., altitude between 250 to 2000 km) that benefit from proximity to earth surface and have been  demonstrated experimentally
\cite{Bedington17:satellite,Liao16:satellite,Yin17:1200km}.

\begin{figure}[t]
    \centering  \includegraphics[width=0.25\textwidth]{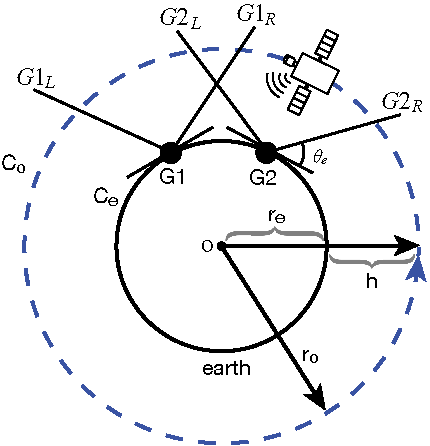}
    \caption{Elevation angle threshold $\theta_{e}$: the elevation angle between the satellite and each ground station must exceed the threshold $\theta_{e}$. In the figure, $r_o$ is the orbit radius, where $r_{e}$ is the earth radius, and $h$ is the satellite altitude.}
    \label{fig:ele-angle}
\end{figure}

To transmit photons successfully to a ground station, the elevation angle, i.e., the angle between the satellite and the horizon at the ground station, needs to exceed a threshold, $\theta_{e}$. For  successful delivery of an entangled pair, the elevation angles between the satellite and the two ground stations must both exceed $\theta_{e}$, as illustrated in Fig.~\ref{fig:ele-angle}. In this figure, the sector between $G_{iL}$ and $G_{iR}$ represent the region where the elevation angle between the satellite and ground station $i$ exceeds $\theta_{e}$, $i=1,2$. The intersection of these two sectors is the region where the satellite can transmit entanglement pairs successfully to both ground stations.

\subsection{Entanglement Sources} \label{sec:SPDC-source}
We assume that the satellite utilizes spontaneous parametric down-conversion (SPDC) based dual-rail polarization entanglement sources that are well-studied and widely used~\cite{Krovi:PDS,Kok00:PDS,dhara2022heralded}. 
In such entanglement sources, a two-qubit entangled Bell state requires four orthogonal modes (i.e., two pairs of mode) to encode. 
The expression of the output is a quantum state as follows\cite{panigrahy2022optimal, dhara2022heralded}:
\begin{dmath}
    \ket{\varphi^{\pm}} = N_{0}\bigg[\sqrt{p(0)}\ket{0,0;0,0}+\sqrt{\frac{p(1)}{2}}(\ket{1,0;0,1}\pm\ket{0,1;1,0})+\sqrt{\frac{p(2)}{3}}(\ket{2,0;0,2}\pm\ket{1,1;1,1}+\ket{0,2;2,0})\bigg],
    \label{eq:spdc_state}
\end{dmath}
where $N_{0}$ is a normalization factor, namely:
\begin{equation}
    N_{0} = \frac{1}{\sqrt{p(0)+p(1)+p(2)}} = \frac{(N_{s}+1)^{2}}{\sqrt{6N_{s}^{2}+4N_{s}+1}}
\end{equation}
and $p(n)$ is the probability of generating a $n$-photon term in each pair of mode, given by:
\begin{equation}\label{eq:p_n}
    p(n) = (n+1)\frac{N_{s}^{n}}{(N_{s}+1)^{n+2}},
\end{equation}
where $N_{s}$ is {\em pump power}, i.e., the mean photon number per mode. The entangled pair from the SPDC dual-rail polarization source is as follows:
\begin{equation}
    \ket{\Psi^{\pm}} = \frac{1}{\sqrt{2}}(\ket{1,0;0,1}\pm\ket{0,1;1,0}),
\end{equation}
the vacuum state is $\ket{0,0;0,0}$, and all the other terms are spurious two-photon states. In Eq. (\ref{eq:spdc_state}), we assume that $N_s$ is low (e.g., below $0.2$) and hence $p(n)$ for $n>2$ is negligible and is omitted in the quantum state. 

The pump power is an important configurable parameter that can be tuned to  maximize the entanglement rate, while adhering to a desired fidelity threshold. In Section~\ref{sec:perf}, we show that the pump power impacts two important factors,  success probability and fidelity, for QKD,  and needs to be chosen carefully.

\subsection{Loss Model}
The satellite quantum communication channel via free space optical (FSO) transmission, must account for the characteristics of the optical channel in its underlying analysis. The transmission losses for each qubit (comprising of a pair of modes) scales  quadratically with free space propagation length and exponentially with aerial propagation length \cite{panigrahy2022optimal}. We incorporate the effect of transmission loss by treating FSO transmission as a Bosonic pure loss channel acting on each mode of the quantum state described in Eq.~\eqref{eq:spdc_state}.
Most generally, the Bosonic pure loss channel leads to a reduction in the mean photon number of the input state; additionally an input pure quantum state becomes a mixed state for non-zero loss. In the present context, this impedes the probability of successfully delivering the entangled pairs to both ground stations, as well as affecting the fidelity (to the ideal Bell state) of the delivered entangled photons  \cite{panigrahy2022optimal}. Namely, it changes the probability of the generation of entangled pairs. The probability of receiving a perfect Bell pair becomes necessarily smaller than $p(1)$ from Eq.~\eqref{eq:p_n}. See more details of the loss model in \cite{panigrahy2022optimal}.

\subsection{Noise Model} \label{sec:noise-model}

Atmospheric FSO transmission channels have to contend with a variety of noise processes. In this manuscript, we limit our noise estimate to unfiltered background photons. Any excess photons in the channel will cause false events (i.e.,  where the qubits of the Bell pair were lost) to be treated as successes, thereby impacting the fidelity of the  entangled pair that should have been delivered. The main contributor of the background photon flux (for example from imperfect filtering by the ground receiver) is commonly associated with the brightness of the sky and varies drastically depending on the time of the day. More specifically, the level of background photon flux is at its highest during clear daylight, and at its lowest during  clear nighttime. In our work, we consider these two setups (i.e., daytime and nighttime) consistent with the state-of-the-art \cite{Harney22} and compute the fidelity of the generated entangled state between two ground stations by modeling the arrival of unfiltered background photons as detector dark click events. 

\section{Blockwise Key Distillation}




We assume the E91 protocol\cite{ekert1991quantum} is used for QKD between the pair of ground stations.  This protocol, like most QKD protocols, consists of a quantum communication stage, followed by a classical post-processing stage.  In the quantum communication stage, $M$ entangled pairs are sent to Alice and Bob (some of which may be lost in transmission due to channel loss).  Alice and Bob then choose, independently at random, whether to measure their particles in the $Z$ or $X$ basis, recording their results.  Later, Alice and Bob will disclose, over the authenticated classical channel, their basis choice, discarding all iterations that do not match.  The resulting strings are called the users' \emph{raw key}. This concludes the quantum communication stage---the output is a raw key of size $N$-bits (with $N \le M$), which may be partially correlated (errors in the channel or adversarial noise may cause errors in Alice and Bob's raw key) and partially secret (Eve may have some non-negligible side information on the raw key based on her attack).  Next, the classical post-processing stage will further process the raw key to produce a secret key.  First the error rate in the raw key is determined.  After this, error correction, and finally privacy amplification protocols are run.  The output of this stage is the final secret key of size $\ell \le N$ bits.  An important metric for the entire QKD protocol is its \emph{key rate}, namely the ratio of the secret key size ($\ell$) to the total number of signals sent ($M$).

\subsection{Blockwise vs. Non-blockwise Schemes}

In this work we analyze and compare two different classical post-processing strategies: \emph{blockwise} and \emph{non-blockwise}.  The latter, non-blockwise, is the traditional QKD scenario whereby the raw key of $N$ bits is treated as a single system from which error correction and privacy amplification are run.  The former, blockwise, divides the raw key up into smaller systems, or blocks.  This division can be arbitrary, but to potentially provide a performance boost, each block should have homogenous channel statistics (especially in terms of noise---that is, while each block may have very different noise levels, the raw keys within a single block should be similar).  In general, if there is a significant difference in the noise levels of the blocks, one can expect blockwise to produce a strictly higher key rate due to the concavity of entropy as discussed below.

First, consider the standard non-blockwise post-processing where all the raw key bits are considered together and is agnostic to the dynamics of the quantum channel. Specifically, in this strategy, a random subset of $m < N/2$ bits is randomly chosen from the set of $N$ bits in the raw key, and Alice and Bob's measurement results are used to estimate the noise of the entire block, denoted as $Q$.  This is defined to be the relative number of bit-flips in Alice and Bob's raw key.  In this work, we assume the noise is modeled by a depolarizing channel and thus (1) the noise is the same in both $Z$ and $X$ bases and (2) we have $Q = (1-F)/2$, where $F$ is the fidelity obtained using the noise model in Section~\ref{sec:noise-model}. 

Under the {\em blockwise} post-processing strategy, 
users break the raw key into blocks of signals based on operating conditions, where each block is expected to have similar noise characteristics.  
In our case, for simplicity, we divide the total raw key into two classes of blocks: one from day and one from night operating conditions as it is expected that the noise in the daylight will be higher than at night. However, blockwise processing can be applied to any arbitrary number of blocks, so long as there are a sufficient number of raw key bits in each block.

More formally, let $rk_A, rk_B\in\{0,1\}^N$ be the raw key for Alice and Bob, and let $B^A_i$ be a single block such that $rk_A = B^A_1B^A_2\cdots B^A_k$ (i.e., $B$ is the bitstring concatenation of each block).  Similarly for $rk_B$ which will have the same decomposition.  Then, a random subset $t_i$ of size $m_i$ is chosen for each block $B_i$ and measurement results in that block, indexed by that subset, are disclosed over the authenticated channel to determine the noise present in each block, denoted $Q_i$.  Finally, error correction and privacy amplification are run on each block separately, distilling $k$ secret keys $s_1$ through $s_k$ which are later concatenated into a single secret key $s$.  The size of each $s_i$ depends on $Q_i$.


\subsection{Key Rate Analysis} \label{sec:analysis}

As the main goal of this paper is to evaluate and compare the effectiveness of both blockwise and non-blockwise key distillation strategies, we require a performance metric.  For this, we will use the \emph{key rate} of the protocol, defined to be the ratio of the number of final secret key bits to the total number of attempted entanglement pairs sent by the source.  Finally, we will also consider both the asymptotic scenario, where the number of signals sent approaches infinity giving us upper-bounds on the key-rates, and the more realistic finite-key scenarios,
where we will also have to take into account imperfect sampling and other imprecisions.  We note that, for this work, we consider idealized photon sources which may emit zero or one photon, but never two.  That is, we set $p(2) = 0$ in Eq.~\ref{eq:spdc_state}.   In our evaluations, $p(2)$ is generally small under the optimal pump power and cannot drastically decrease the key rate, as we shall show in  Section~\ref{sec:multi-photon}.  However, a rigorous blockwise and non-blockwise analysis for multiphoton sources remains an interesting future challenge.

To compute the key rate of the protocol, both in the blockwise and non-blockwise cases, we turn to analysis methods derived in \cite{tomamichel2012tight} which utilize entropic uncertainty \cite{tomamichel2011uncertainty}.  

\smallskip
\noindent{{\bf Key rate analysis for non-blockwise scheme.}} First consider the non-blockwise case.  Here, the entire raw-key is treated as a single system from which a random sample of size $m$ is chosen (leaving $n=N-m$ bits for the raw key).  This sample allows parties to estimate the error in the entire raw key denoted as $Q$.  From this, the remaining signals are run through an error correction process (leaking an additional $\lambda_{EC}$ bits to the adversary). A  test is then run by hashing the error corrected raw key and testing correctness between Alice and Bob (which leaks $\log\frac{1}{\epsilon_{cor}}$ bits to the adversary for user specified $\epsilon_{cor}$).  Finally, privacy amplification is run, outputting a secret key of size $\ell \le n$. It is guaranteed that, conditioning on not aborting the protocol, the final secret key system and Eve's ancilla (denoted as $\rho_{KE}$) will satisfy the following:
\[
\frac{1}{2}\trd{\rho_{KE} - I/2^\ell \otimes\rho_E} \le \epsilon_{sec},
\]
where $\rho_E$ is Eve's system.
That is, the final secret key system will be $\epsilon_{sec}$ close (in trace distance) to a truly uniform random key, $I/2^\ell$, which is also completely independent of Eve's system $\rho_E$.

Using results in \cite{tomamichel2012tight}, the non-blockwise case can be shown to have an overall secret key length of:
\begin{equation} \label{eq:keyrate-nonblock}
    \ell_{non-block} = n(1-h(Q+\mu)) - \lambda_{EC} - \log\frac{2}{\epsilon_{sec}^2\epsilon_{cor}}.
\end{equation}
where $\epsilon_{cor}$ is a security parameter determining the failure rate of the correctness portion of the protocol (i.e., Alice and Bob will have the same secret key, except with probability at most $\epsilon_{cor}$), and $\epsilon_{sec}$ is a security parameter determining the distance of the final secret key from a truly uniform random key.  Above, $\lambda_{EC}$ represents the information leaked during error correction; in our evaluations, later, we simply set $\lambda_{EC} = nh(Q+\mu)$.  Other realistic settings of $\lambda_{EC}(Q) = 1.2nh(Q)$ can also be used. However such a setting will not significantly affect our results in later sections as we are primarily interested in comparing blockwise to non-blockwise.  Finally, $\mu$ is a result of finite sampling effects and is set to:
\[
\mu = \sqrt{\frac{(n+m)(m+1)}{nm^2}\ln\frac{2}{\epsilon_{sec}}}.
\]
The above may be derived from standard classical sampling arguments \cite{tomamichel2012tight}.

\smallskip
\noindent{{\bf Key rate analysis for blockwise scheme.}} In the blockwise case, the setting is similar and we may again use results from \cite{tomamichel2012tight} to distill each sub-block into secret keys independently, and then concatenate the final blockwise secret keys into a single secret key.  Here, let $B_i$ be the size of the $i$'th block (determined by the user).  Now, a random subset $t_i$ of size $m_i$ for each block $B_i$ is chosen.  As with the non-blockwise, this sampling subset is used to determine the error rate in the raw key, however, now, it is used only to estimate the error rate in the $i$'th block of the raw key, denoted $Q_i$.  Error correction, a correctness test, and finally privacy amplification is then performed individually on each block.  From this setup, we can compute the secret key size of block $i$ to be:
\[
\ell_i = (B_i-m_i)(1-h(Q_i+\mu_i)) - \lambda_{EC}^{(i)} - \log\frac{2}{\epsilon_{sec}^2\epsilon_{cor}},
\]
from which we have the following total secret key size:
\begin{equation} \label{eq:keyrate-block}
    \ell_{block} = \sum_{i=1}^kn_i(1-h(Q_i+\mu_i)) - \sum_i\lambda_{EC}^{(i)} - k\log\frac{2}{\epsilon_{sec}^2\epsilon_{cor}},
\end{equation}
where $k$ is the total number of blocks and $n_i = B_i-m_i$.  The value of $\mu_i$ is identical to $\mu$ above, except replacing $m$ with $m_i$ and $n$ with $B_i-m_i$.  Finally, $\lambda_{EC}^{(i)}$ is the amount of information leaked during error correction of block $i$.  In our evaluations, we set this to $\lambda_{EC}^{(i)} = (B_i-m_i)h(Q_i+\mu_i)$ .

The above values of $\ell$ can be used to immediately compute the key rate simply by dividing by the number of attempted entanglement pairs sent by the satellite (we say attempted as the satellite may send out vacuum states which count, detrimentally, to the overall key rate).

To determine theoretical upper-bounds, we also consider the asymptotic scenario, where the number of signals approaches infinity.  In this instance,  the key rate for the non-blockwise scheme is simply $1-2h(Q)$, while the key rate for the blockwise scheme converges to $\sum_i p_i(1-2h(Q_i))$, where $p_i$ is the proportion of total raw keybits used in block $i$ as the size of the raw key approaches infinity.

Note that the above equations give immediate intuition as to why blockwise processing can lead to higher key-rates.  For non-blockwise, the total error $Q$ is actually the average error over all individual blocks.  Due to the concavity of Shannon entropy, the key rate can only be higher, but no less than, the blockwise processing, \emph{at least in the asymptotic scenario.}  In the finite key scenario, sampling imprecisions lead to other problems and, so, as we show later blockwise processing can actually lead to worse results in some settings.  Knowing when to use blockwise processing and when to use non-blockwise is an important question to answer if these systems are to be practically deployed.

\section{Performance Evaluation} \label{sec:perf}

In this section, we evaluate the performance of blockwise and non-blockwise key distillation schemes. 
As mentioned in Section~\ref{sec:sate-QKD},  we consider a LEO satellite, with two ground stations on the equator. In the following, we first describe the evaluation setup and then the results. 

\subsection{Evaluation Setup} 
We consider  three satellite altitudes,  $A=500$ km, 800 km, 1000 km. For each satellite altitude, we consider two ground stations along the equator of the Earth, with a distance of $D=600$ km, 1200 km, or 1800 km. The satellite is equipped with a SPDC entanglement source (see Section~\ref{sec:SPDC-source}) that operates at a 1 GHz rate, i.e., generating $10^9$ entangled photons per second. The elevation angle threshold (Section~\ref{sec:sate-QKD}) is set to 20$^{\circ}$. For simplicity, we assume 10 hours of nighttime (8pm-6am), and the remaining 14 hours as daytime each day.  The dark click probability 
$P_d$ is set to $3 \times 10^{-6}$ for nighttime and $3 \times 10^{-3}$ for daytime based on the study in~\cite{pirandola2021satellite}.  
The blockwise scheme treats the raw key bits produced during daytime and nighttime separately, i.e., it considers two types of blocks, corresponding to the raw keys from daytime and nighttime respectively, while the non-blockwise scheme considers all of the raw key bits together. 

When the satellite altitude is 500 km, the orbit time (the amount of time for a satellite to finish one orbit) is 5,647 seconds, for satellite altitudes of 800 and 1000 km, the orbit time is longer (6,022 and 6,276 seconds, respectively). The number of passes of the satellite over the two ground stations is 6 passes during nighttime and 9 passes during daytime for all the settings, except when the satellite altitude is 500 km, where the number of passes during nighttime is 7. 
Table~\ref{table:contact_length} lists the contact length (i.e., pass duration, the duration that the satellite is in the contact of both ground stations) for the various settings. 
The contact length varies from less than 1 minute to over 8 minutes. As expected, for a given satellite altitude, larger ground station distance leads to shorter contact length; while for the same ground station distance, higher satellite altitude leads to longer contact length. Using the loss and noise models in Section~\ref{sec:prelim}, we obtain the success probability and fidelity of the transmission from the satellite to each ground station in each second, and then obtain the average success probability and fidelity over the contact length in the following evaluation. 

We consider the key rate for running the protocol over 1 to 80 days to show the performance of the two key distillation schemes over time as more raw key bits are accumulated at the ground stations, and the performance of these schemes relative to the asymptotic results. 
For both blockwise and non-blockwise schemes, we vary the pump power 
of the SPDC source (see Section~\ref{sec:SPDC-source}) and the sampling rate for each setting so that the number of secret key bits is maximized. Specifically, the pump power is varied from 0 to 0.1.  We limit the pump power up to 0.1 so that the approximation of the quantum states in \S\ref{sec:SPDC-source} is accurate and  high-order-photon contributions are negligible~\cite{Dhara21:dual-rail-PDS}. 
The sampling rate is varied from $5\times 10^{-4}/k$ to $3\times 10^{-1}/k$ for the raw keys generated in $k$ days. 

Unless otherwise stated, our results below assume that $p(2)$, i.e., the probability of generating a 2-photon term in each pair of mode of the SPDC source, is zero 
(see Section~\ref{sec:analysis}). In Section \ref{sec:multi-photon}, we show that this is a reasonable approximation.

\begin{table}[t]
\centering
\caption{Satellite contact length (seconds) with ground stations.}

\begin{tabular}{ |c||c|c|c|  }
 \hline
 & $D=600$ km & $D=1200$ km & $D=1800$ km\\
 \hline \hline
$A=500$ km & 224  & 134 &  43 \\
 \hline
$A=800$ km & 385  & 288  &   190 \\
\hline
$A=1000$ km  & 488     & 387 &   285 \\
\hline
\end{tabular}
\label{table:contact_length}
\end{table}

\subsection{Impact of Pump Power}



\begin{figure}[t]
	\centering
	\hspace*{-.5cm}
    \subfigure[Success prob., night]
    {\includegraphics[width=0.240\textwidth]{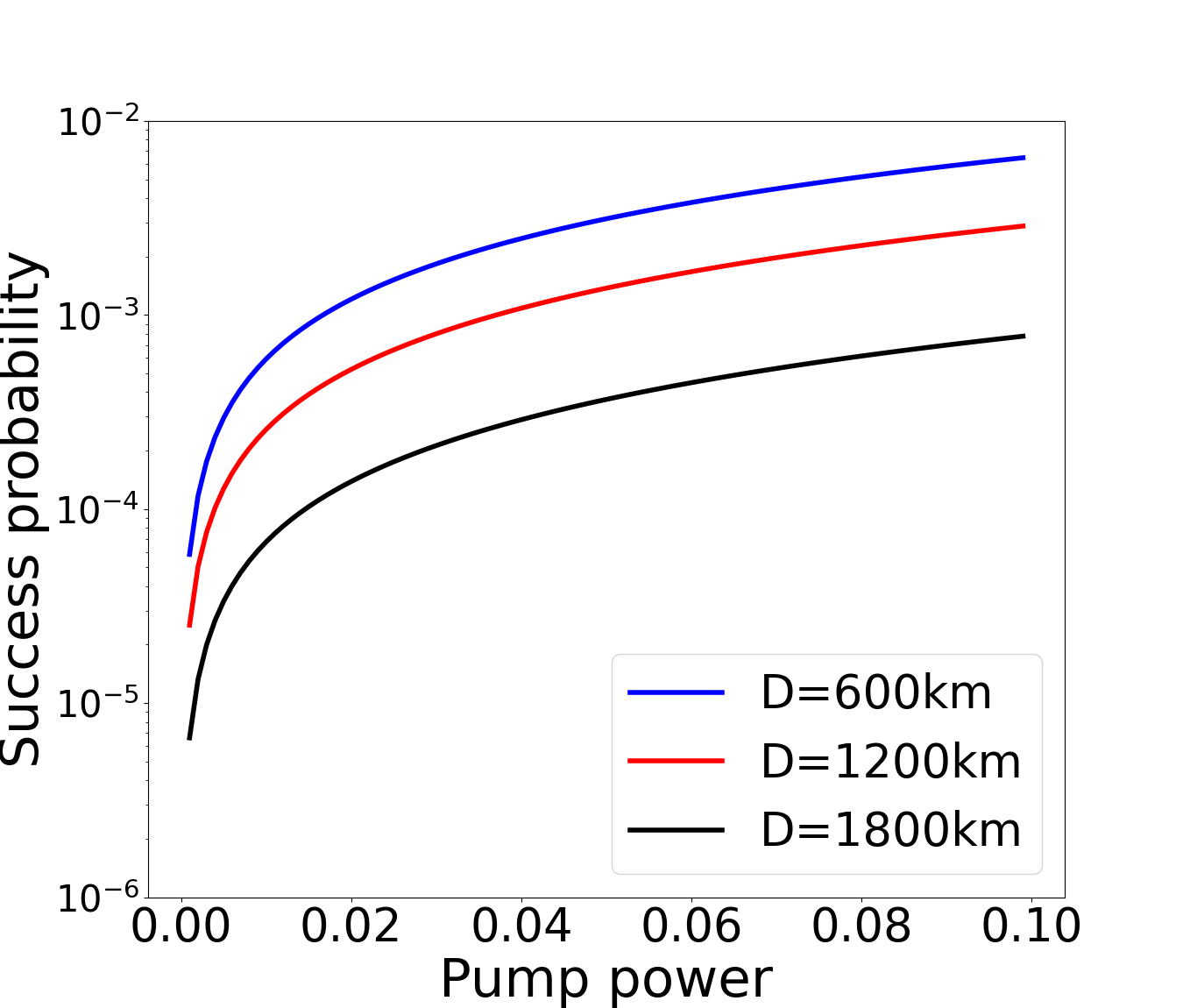}}
	\hspace*{-.5cm}
    \subfigure[Fidelity, night]
    {\includegraphics[width=0.240\textwidth]{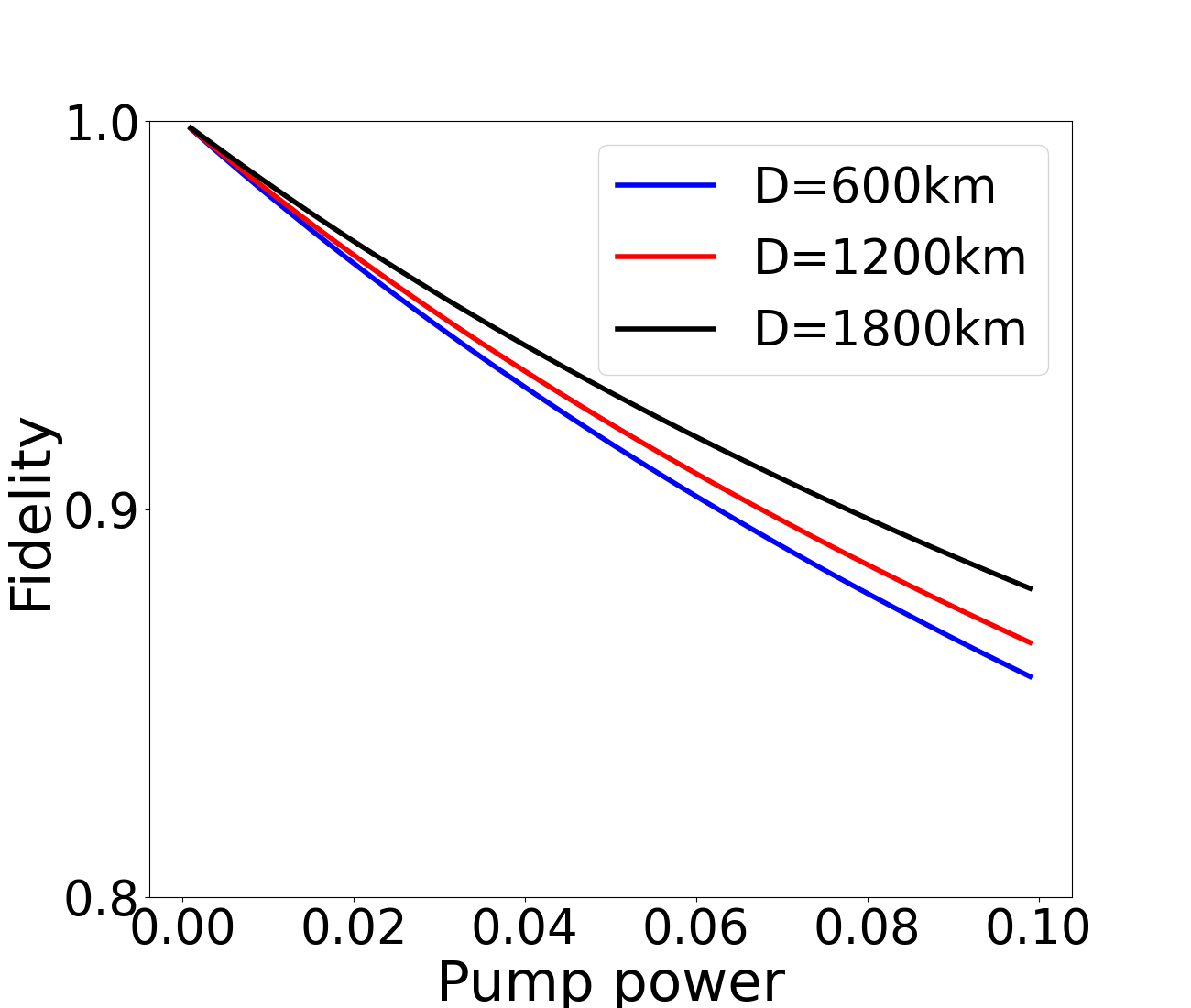}}
	\hspace*{-.5cm}
    \subfigure[Success prob., day]
    {\includegraphics[width=0.240\textwidth]{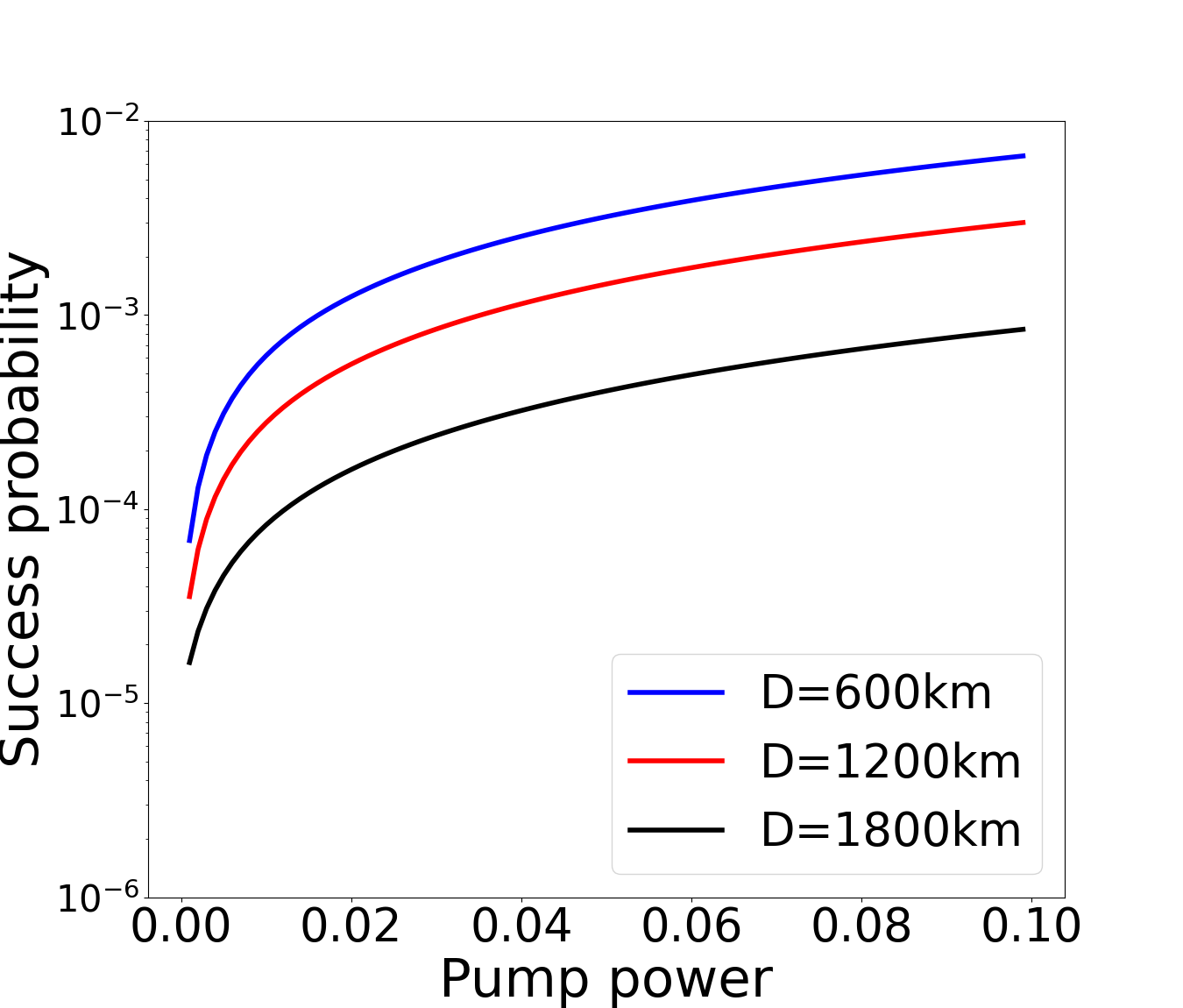}}
	\hspace*{-.5cm}
    \subfigure[Fidelity, day] 
    {\includegraphics[width=0.240\textwidth]{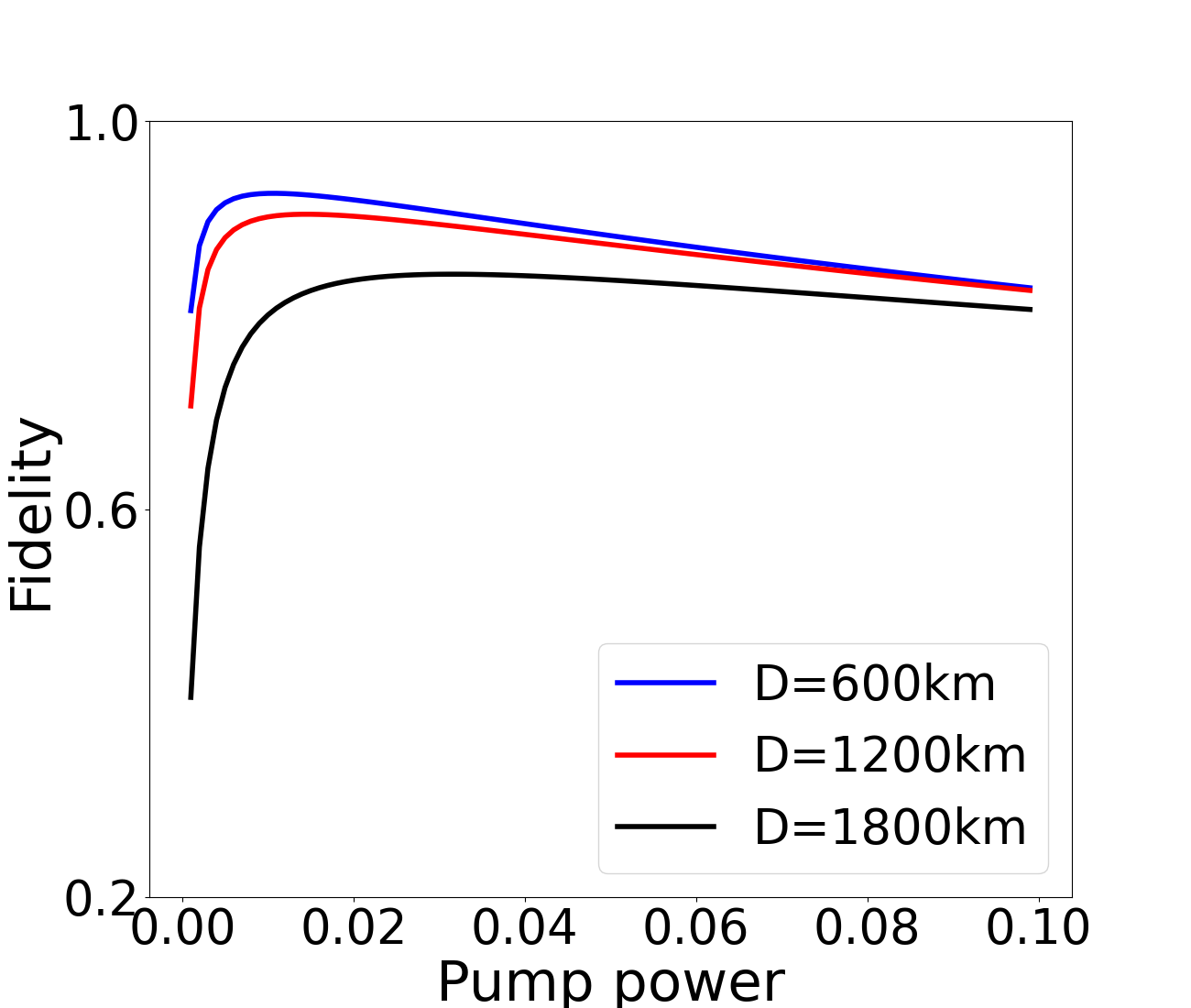}}
	\vspace{-.10in}
	\caption{ {\small Impact of pump power on success probability and fidelity, satellite altitude $A=500$ km. } 
    } 
	\label{fig:pump-power-500km}
	\vspace{-.1in}
\end{figure}

\begin{figure*}[ht]
    \centering
    \subfigure[Optimal pump power]{\includegraphics[width=0.25\textwidth]{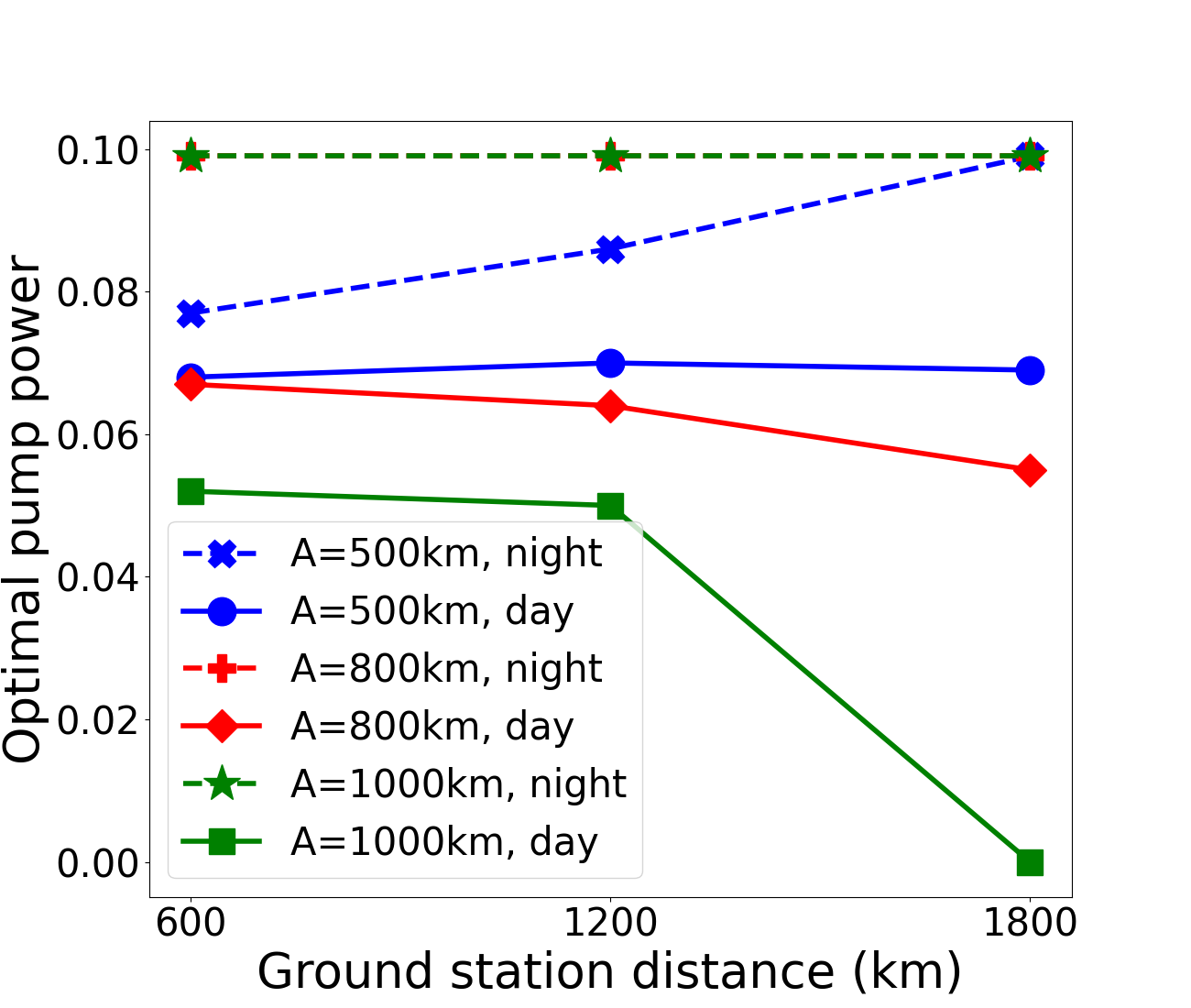}}
    \hspace{-.5cm}
    \subfigure[Success prob.]{\includegraphics[width=0.25\textwidth]{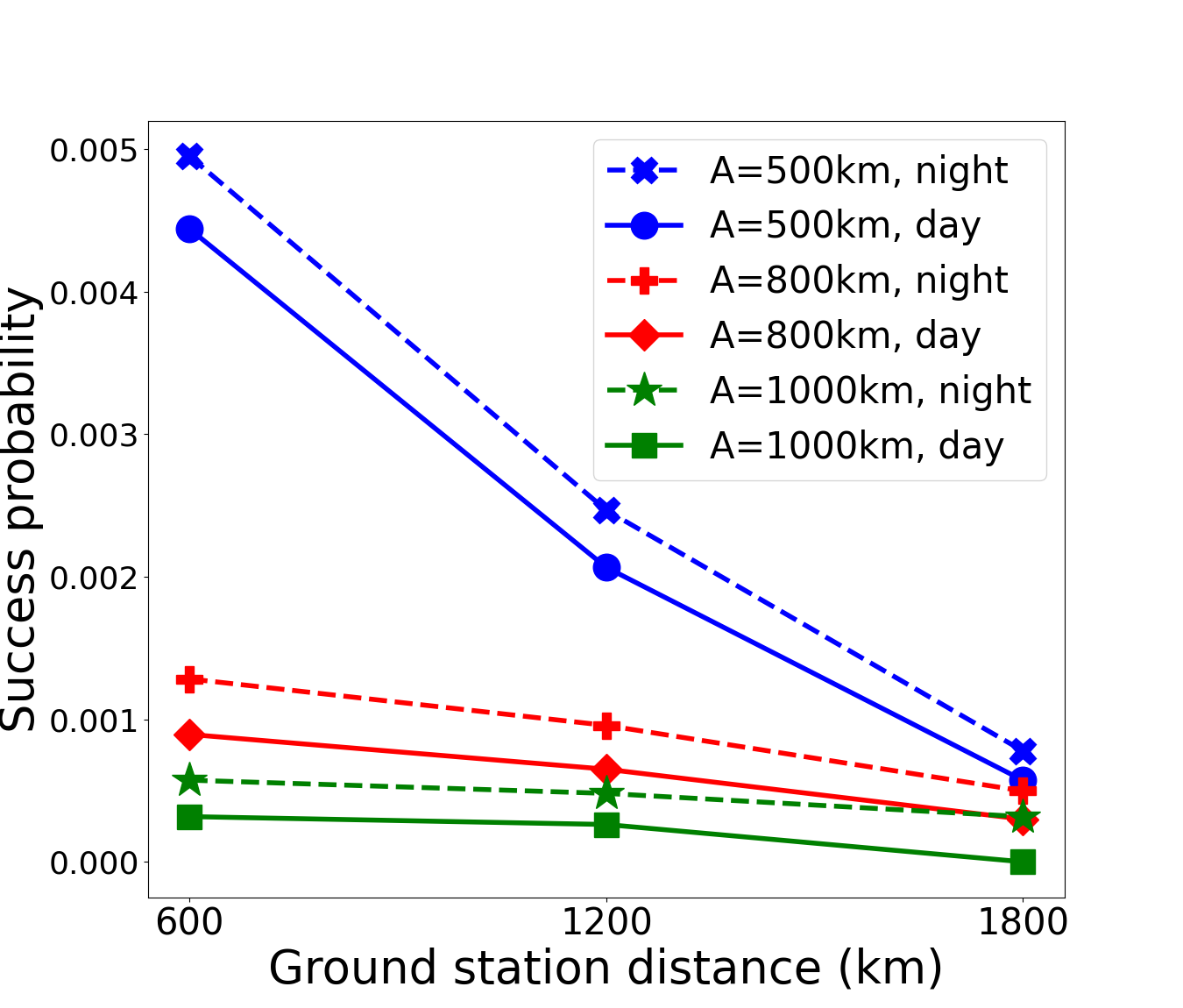}} 
    \hspace{-.5cm}
    \subfigure[Fidelity]{\includegraphics[width=0.25\textwidth]{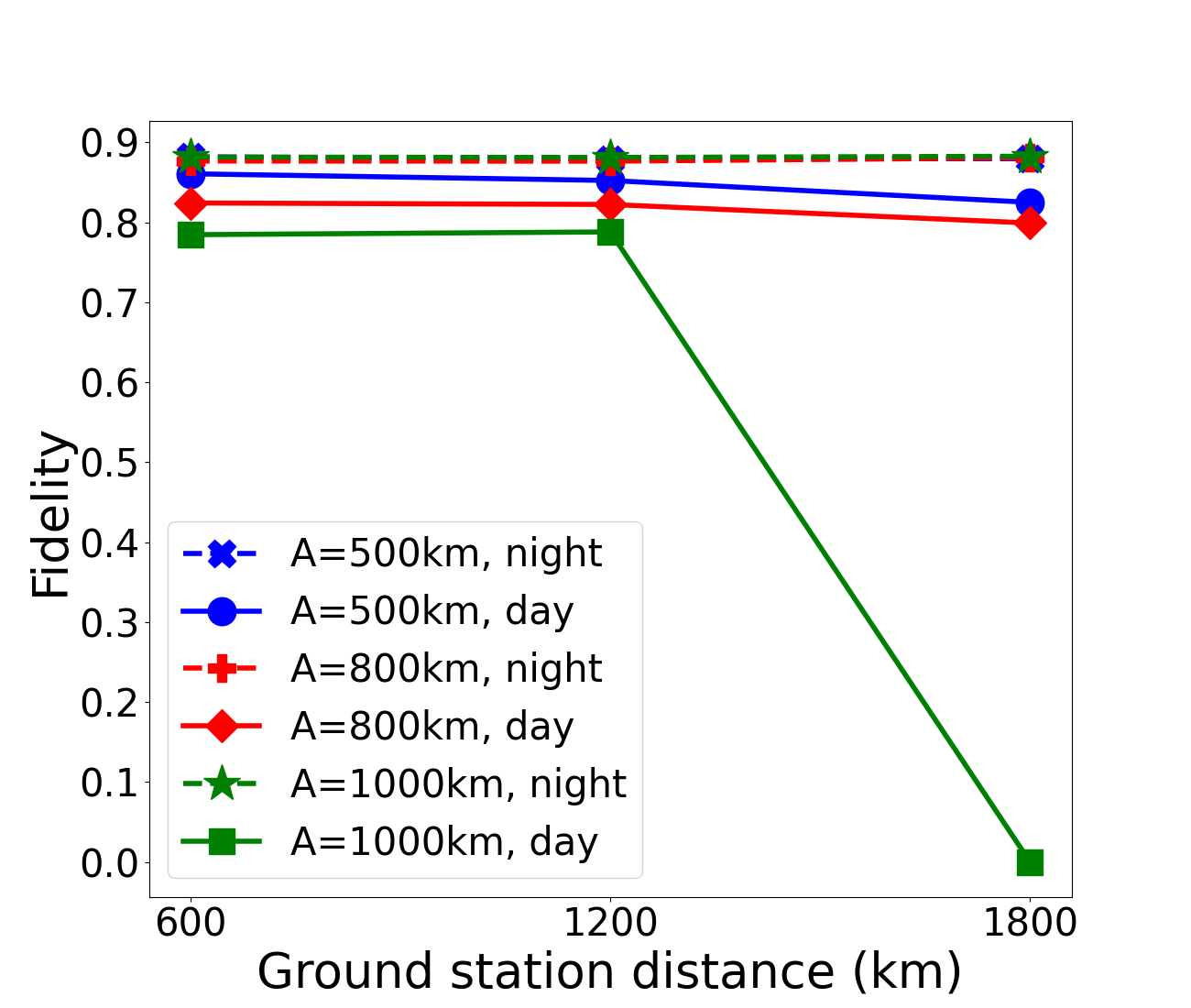}}
    \hspace{-.5cm}
    \subfigure[Optimal sampling rate]{\includegraphics[width=0.25\textwidth]{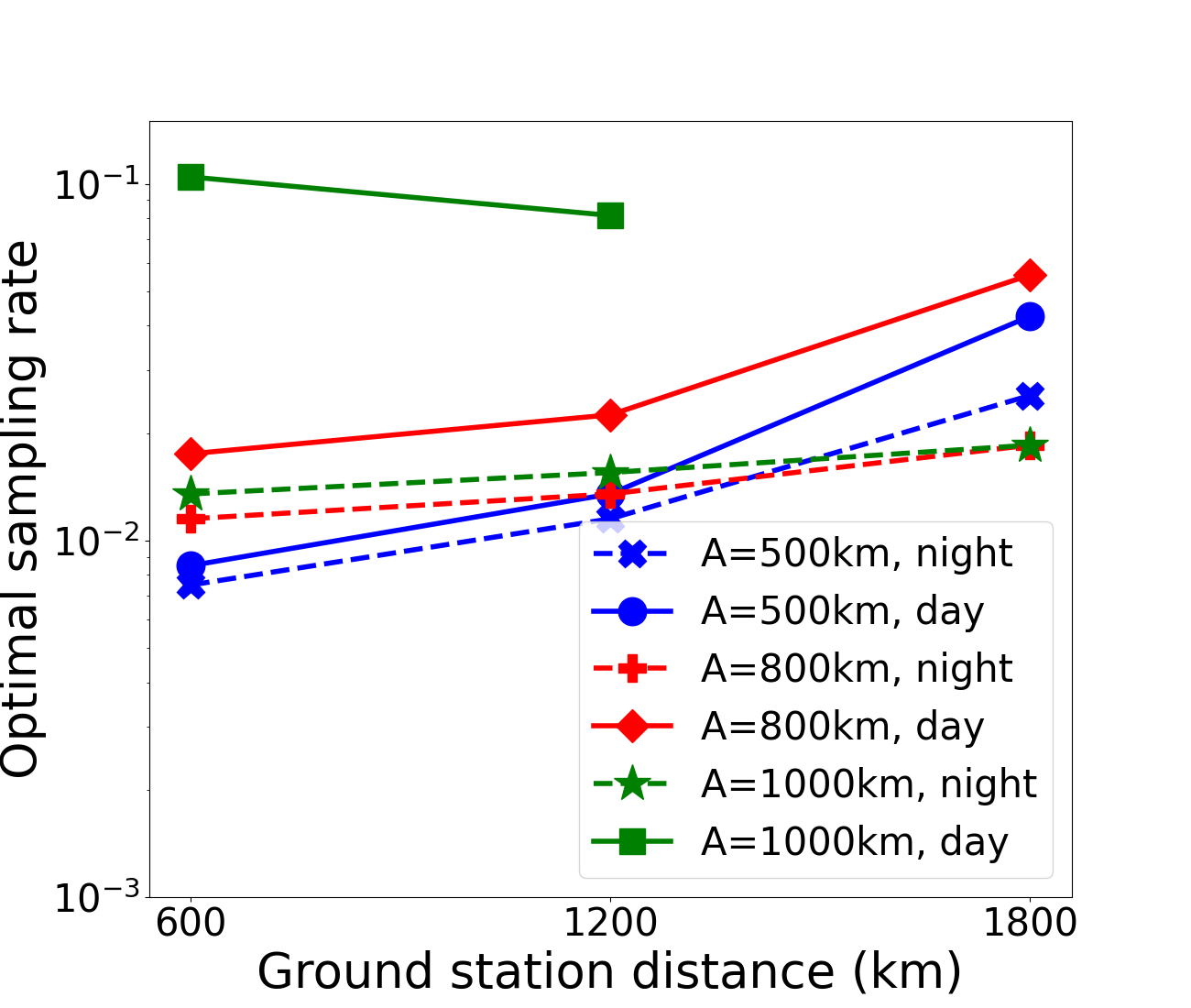}}
    \hspace{-.5cm}
    \caption{{\small Blockwise scheme: optimal pump power, the resultant success probability and fidelity, and optimal sampling rate (1 day).}  }
    \label{fig:opt-pumppower-samplingrate-blockwise}
\end{figure*}

\singlefig{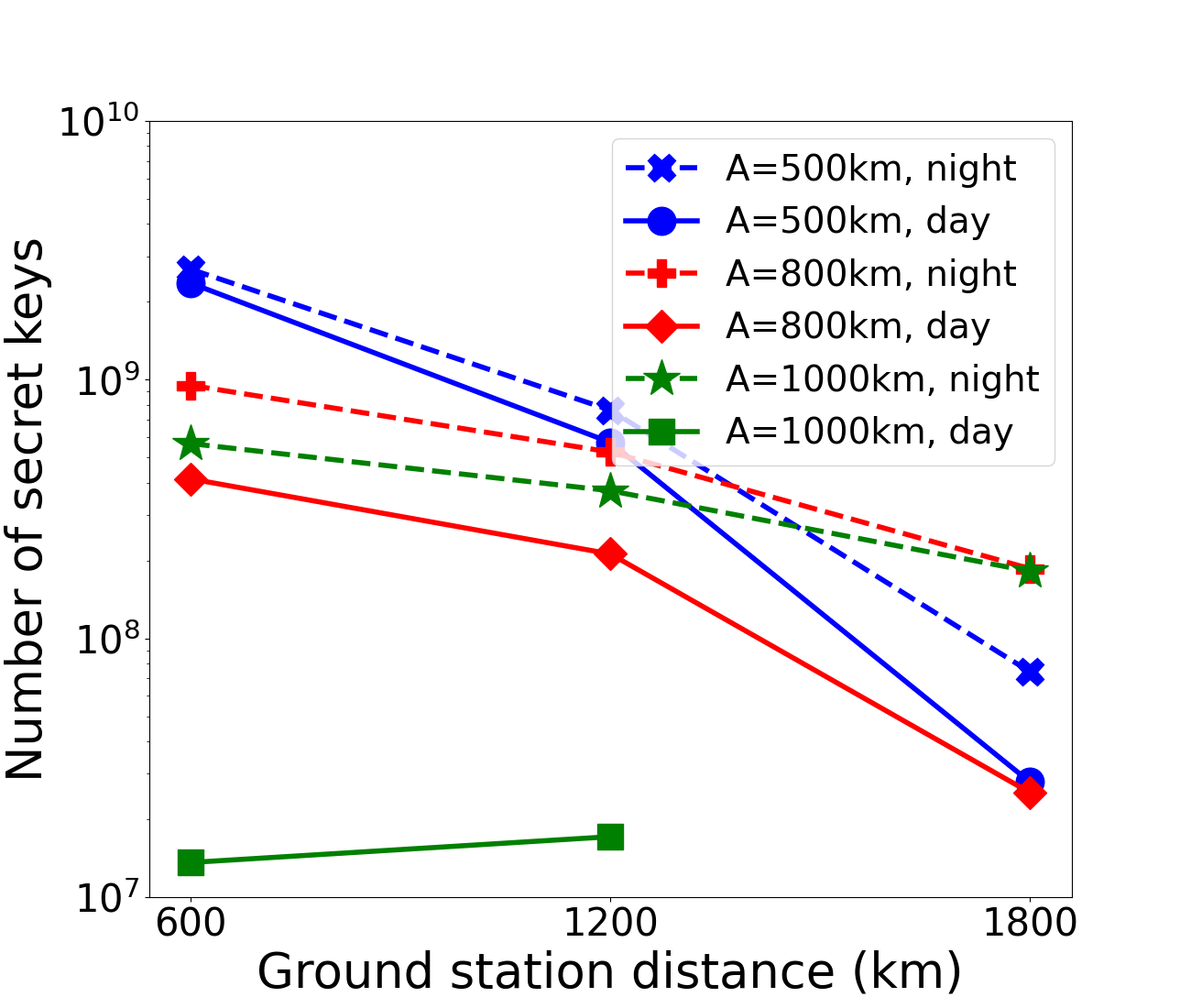}{ Blockwise scheme: Number of secret key bits that are generated under the optimal pump power and sampling rate (1 day). }{num-key-blockwise}

We first examine the impact of the pump power on the success probability and fidelity in the various settings.  
Fig.~\ref{fig:pump-power-500km} plots the success probability and fidelity as a function of pump power when the satellite altitude $A=500$ km. Figures~\ref{fig:pump-power-500km}(a) and (b) show the results for nighttime, where the results for various ground station distances are shown in the figure. We see that, for all three  ground station distances, success probability increases with pump power, while fidelity decreases with pump power. In addition, for the same pump power, a shorter ground station distance leads to a larger success probability, but lower fidelity.    
Figures~\ref{fig:pump-power-500km}(c) and (d) show the results for daytime. In this case, while we see similar trend for success probability as that for nighttime, the relationship between fidelity and pump power is more complex: fidelity first increases and then decreases with the pump power. In addition, for the same pump power, while a shorter ground station distance again leads to higher success probability as that in nighttime, it leads to higher fidelity in daytime, opposite to the observation in nighttime.    


Results for the other two satellite altitudes (800 and 1000 km) show similar trends, with variations in the relative relationship among the three ground station distances. For instance, when $A=1000$km, for the same pump power, the fidelity for the three ground station distances is very close to each other during nighttime, while the fidelity for $D=1200$ km is larger than that for $D=600$ km, followed by that of $D=1800$ km.

\begin{figure*}[ht]
    \centering
    \subfigure[Optimal pump power]{\includegraphics[width=0.25\textwidth]{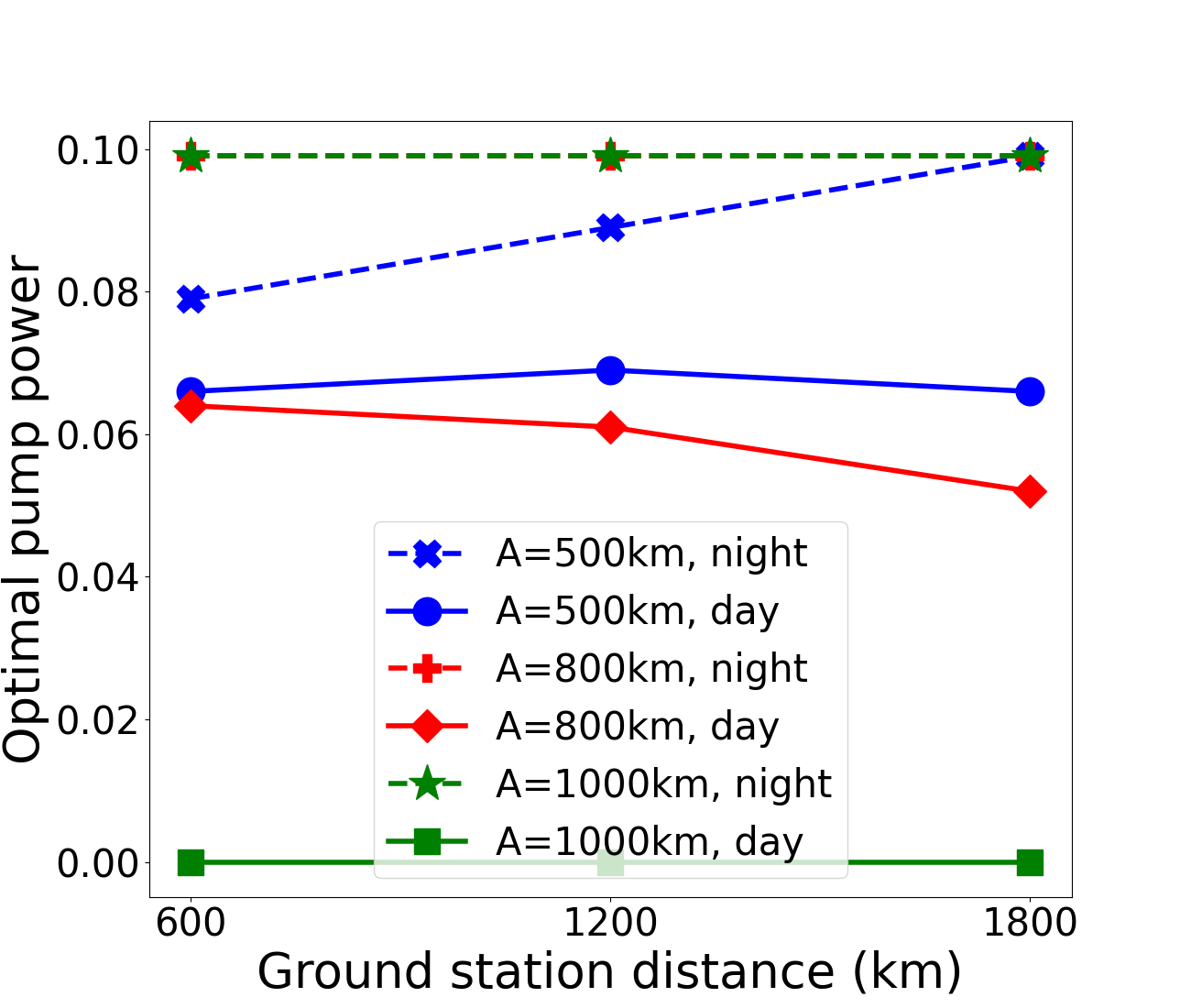}}
    \hspace{-.5cm}
    \subfigure[Success prob.]{\includegraphics[width=0.25\textwidth]{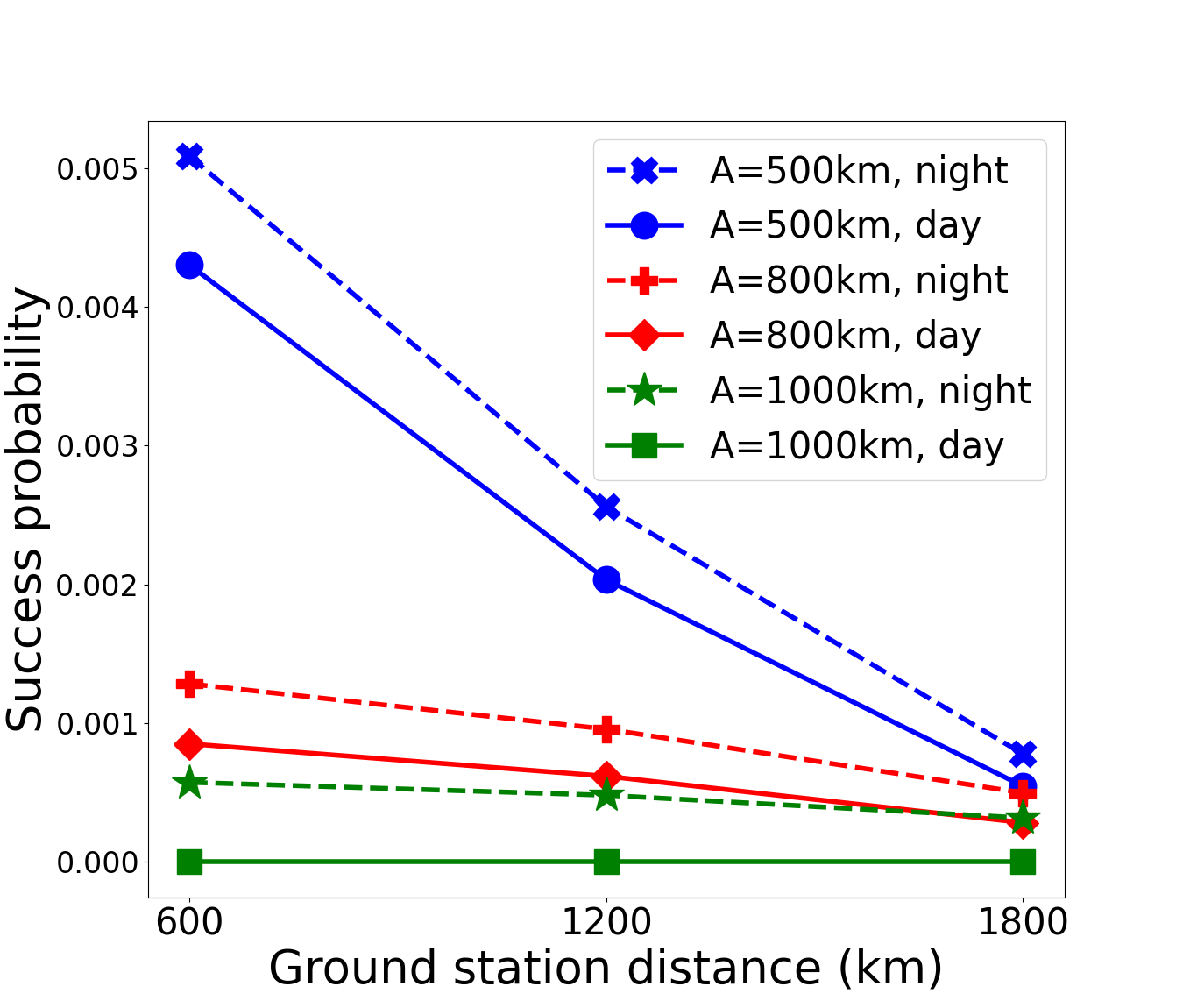}} 
    \hspace{-.5cm}
    \subfigure[Fidelity]{\includegraphics[width=0.25\textwidth]{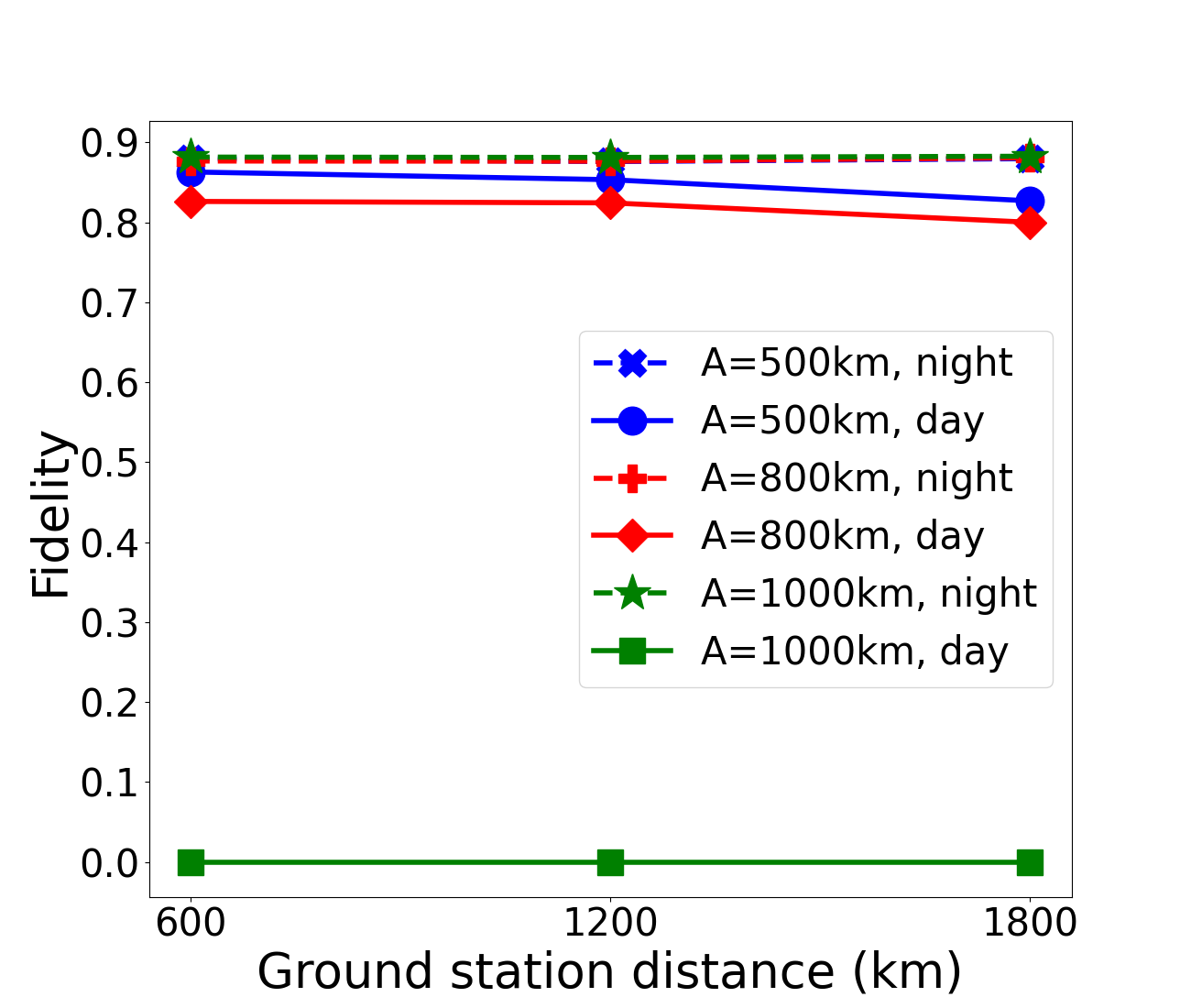}}
    \hspace{-.5cm}
    \subfigure[Optimal sampling rate]{\includegraphics[width=0.25\textwidth]{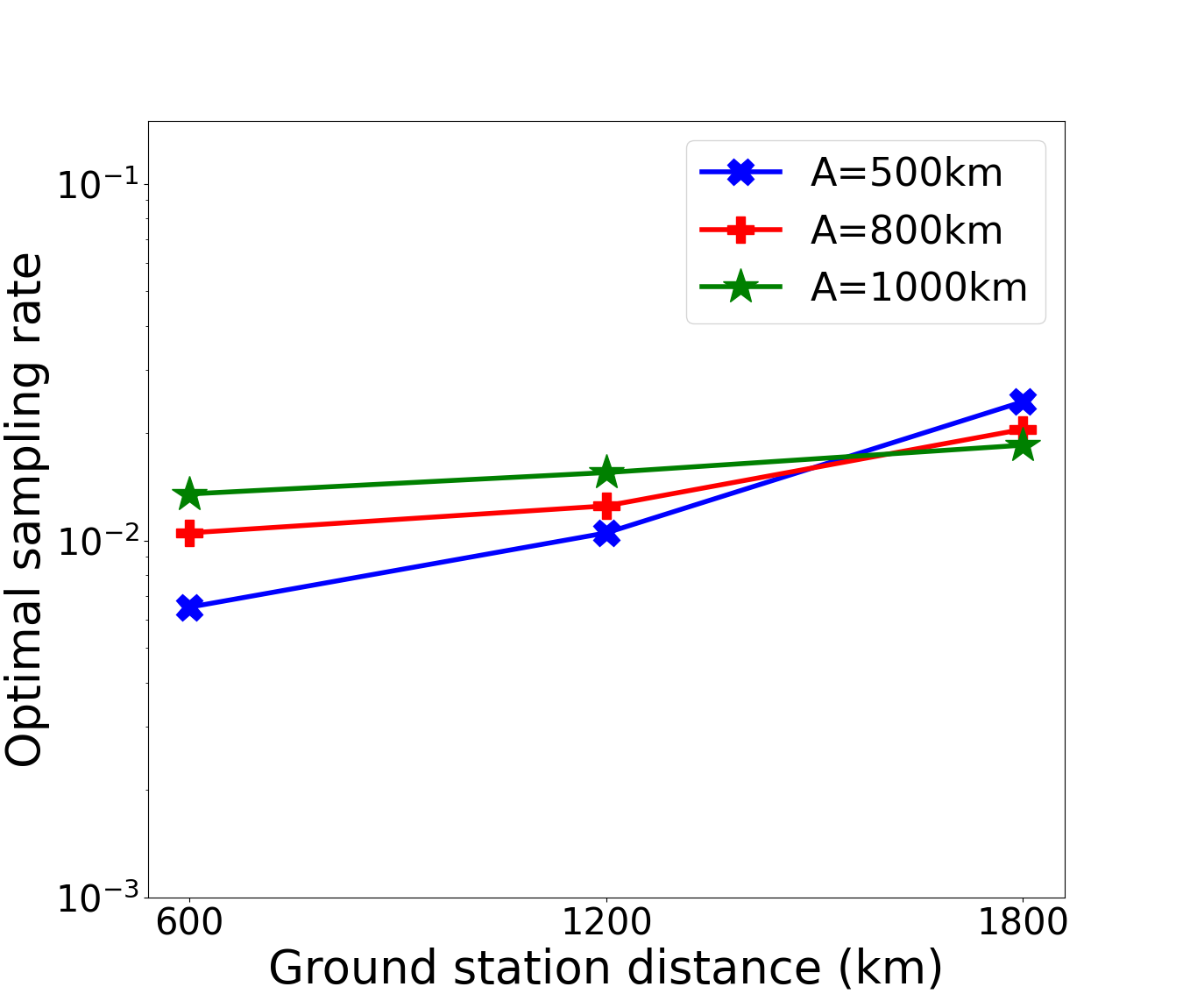}}
    \hspace{-.5cm}
    \caption{Non-blockwise scheme: optimal pump power, the resultant success probability and fidelity, and optimal sampling rate (1 day). }
    \label{fig:opt-pumppower-samplingrate-non-blockwise}
\end{figure*}

\singlefig{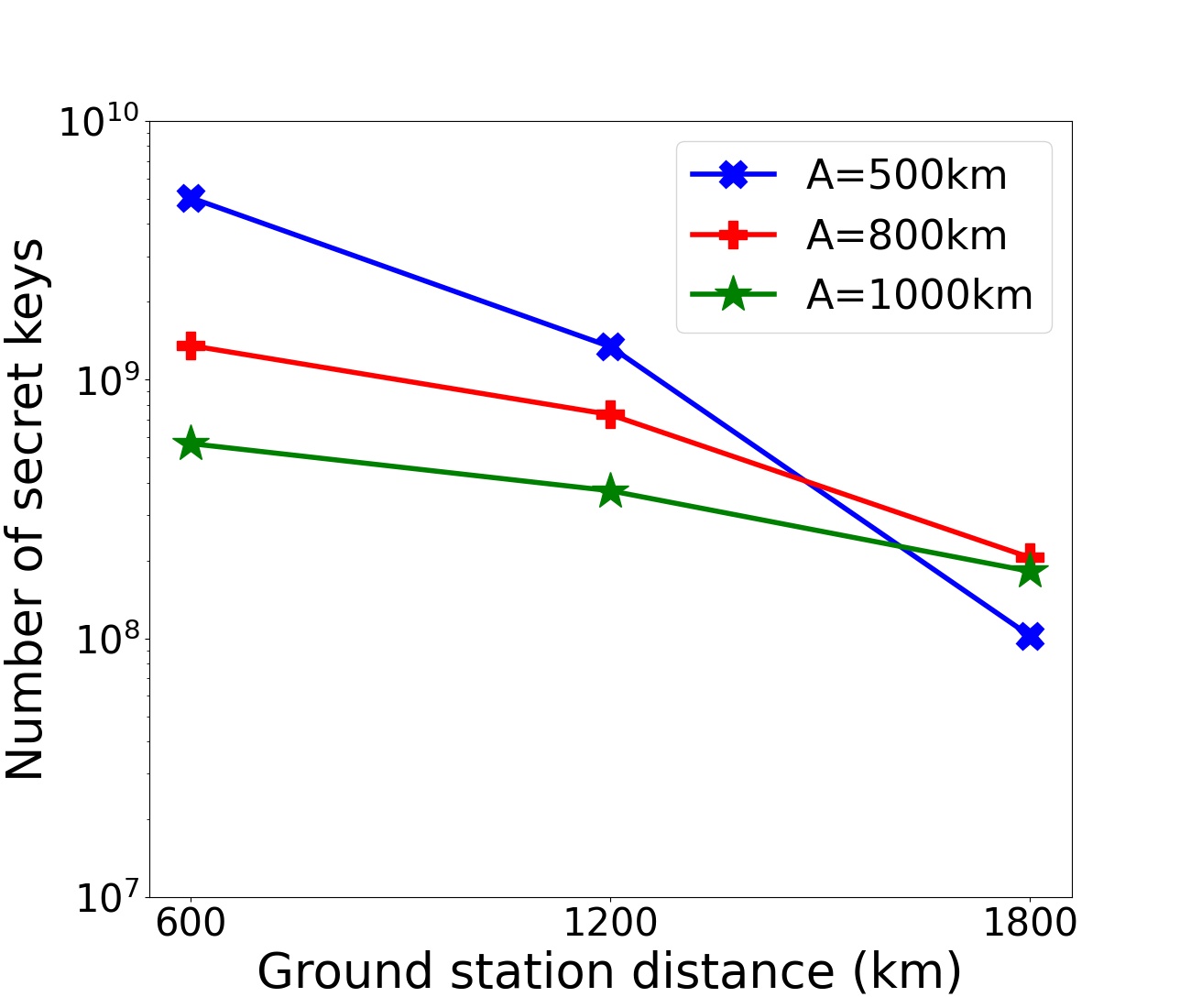}{Non-blockwise scheme: Number of secret key bits that are generated under the optimal pump power and sampling rate (1 day). }{num-key-non-blockwise}

\subsection{Optimal Pump Power and Sampling Rate} \label{sec:opt-pump-power}

Since secret key rate is affected by both success probability and  fidelity, while these two factors are affected by pump power in roughly opposite ways as shown above, we need to find the optimal pump power to maximize the secret key generation rate for various settings. The optimal pump power hence may differ, depending on  satellite altitude, ground station distance, nighttime versus daytime, and also the key distillation scheme. In addition, as shown in Eq.~(\ref{eq:keyrate-nonblock}) and Eq.~(\ref{eq:keyrate-block}),  secret key rate is also affected by sampling rate. In the following, we show the optimal pump power and sampling rate for the various settings for key generation in one day; the results for multiple days are deferred to Section \ref{sec:multi-days}.

\subsubsection{Blockwise Scheme}
Fig.~\ref{fig:opt-pumppower-samplingrate-blockwise} plots results for the blockwise distillation scheme. Specifically, Fig.~\ref{fig:opt-pumppower-samplingrate-blockwise}(a) shows the optimal pump power for various combinations of satellite altitude and ground station distance; results for both nighttime and daytime are plotted in the figure. For the same satellite altitude and ground station distance, we see the optimal pump power for nighttime is larger than that for daytime. Specifically, for satellite altitude of 800 and 1000 km, the optimal pump power for nighttime is 0.1, the maximum pump power that is allowed, and for satellite altitude of 500 km, the optimal pump power is close or equal to 0.1.  For daytime, under the same ground station distance, the optimal pump power is lower for higher satellite altitude. As a special case, when the ground station distance is 1800 km, the optimal pump power for daytime is 0 when the satellite altitude is 1000 km, since no key can be generated for any values of the range of pump power.   

Figures~\ref{fig:opt-pumppower-samplingrate-blockwise}(b) and (c) plot the success probability and fidelity under the optimal pump power for the various settings. For the same satellite altitude and ground station distance, the success probability and fidelity for nighttime are both larger than their corresponding values for daytime. In addition, for the same ground station distance, the success probability under the optimal pump power for lower satellite altitude tends to be larger, for both nighttime and daytime. For fidelity, the optimal fidelity is similar for all nighttime settings, while for daytime, lower satellite altitude tends to have higher fidelity for the same ground station distance.   When the ground station distance is 1800 km and the satellite altitude is 1000 km, both the success probability and fidelity are 0, since the optimal pump power for that case is 0. 

Fig.~\ref{fig:opt-pumppower-samplingrate-blockwise}(d) plots the optimal sampling rate for the various settings. The optimal sampling rate varies from 0.0075 to 0.1045, with higher optimal sampling rate for daytime than nighttime under the same satellite altitude and ground station distance. For a given satellite altitude, larger ground station distances tend to require higher optimal sampling rates.

Fig.~\ref{fig:num-key-blockwise} plots the number of secret key bits generated over daytime and nighttime in a day for the various settings. For each satellite altitude, the number of secret keys generated decreases with ground station distance for both nighttime and daytime, except for a satellite altitude of 1000 km during daytime. For the same ground station distance, lower satellite altitude tends to lead to more secret keys, except for one case (satellite altitude of 500 km and  ground station distance of 1800km), which leads to fewer key bits than the satellite altitude of 800 km due to the significantly shorter contact length in this scenario than others (see Table~\ref{table:contact_length}).    

\subsubsection{Non-blockwise Scheme}


Fig.~\ref{fig:opt-pumppower-samplingrate-non-blockwise} plots the results for the non-blockwise scheme. For various settings, the optimal pump power in Fig.~\ref{fig:opt-pumppower-samplingrate-non-blockwise}(a) is similar to that under the blockwise scheme, except that when the satellite altitude $A=1000$ km, the optimal pump power for daytime is 0 for all ground station distances. This is because the fidelity in the daytime is low for all the pump power values, which leads to higher average error rate (across nighttime and daytime), and overall lower number of keys, compared to the case when only generating keys at nighttime.  Fig.~\ref{fig:opt-pumppower-samplingrate-non-blockwise}(b) and (c) plot the resultant success probability and fidelity for the various settings with the optimal pump power. They are similar to those for the blockwise scheme except when the satellite altitude $A=1000$ km and daytime. Fig.~\ref{fig:opt-pumppower-samplingrate-non-blockwise}(d) plots the optimal sampling rate, which is in the range of $0.0065$ and $0.0245$. Similar to that of the blockwise case, for a given satellite altitude, larger ground station distances have higher optimal sampling rates. 

Fig.~\ref{fig:num-key-non-blockwise} plots the number of secret key bits generated using the non-blockwise scheme in a day for the various settings. Since this scheme combines the raw keys generated during nighttime and daytime together, we simply plot the overall number of keys over a day. For the same ground station distance, more keys are generated at lower satellite altitude, except for one case,  
the satellite altitude is 500 km and the ground station distance is 1000 km, due to its significantly shorter contact length than other settings (see Table~\ref{table:contact_length}).  




\subsection{Compring Blockwise and Non-blockwise Schemes}
\label{sec:multi-days}

\begin{figure*}[ht]
    \centering
    \subfigure[$A=500$ km]{\includegraphics[width=0.33\textwidth]{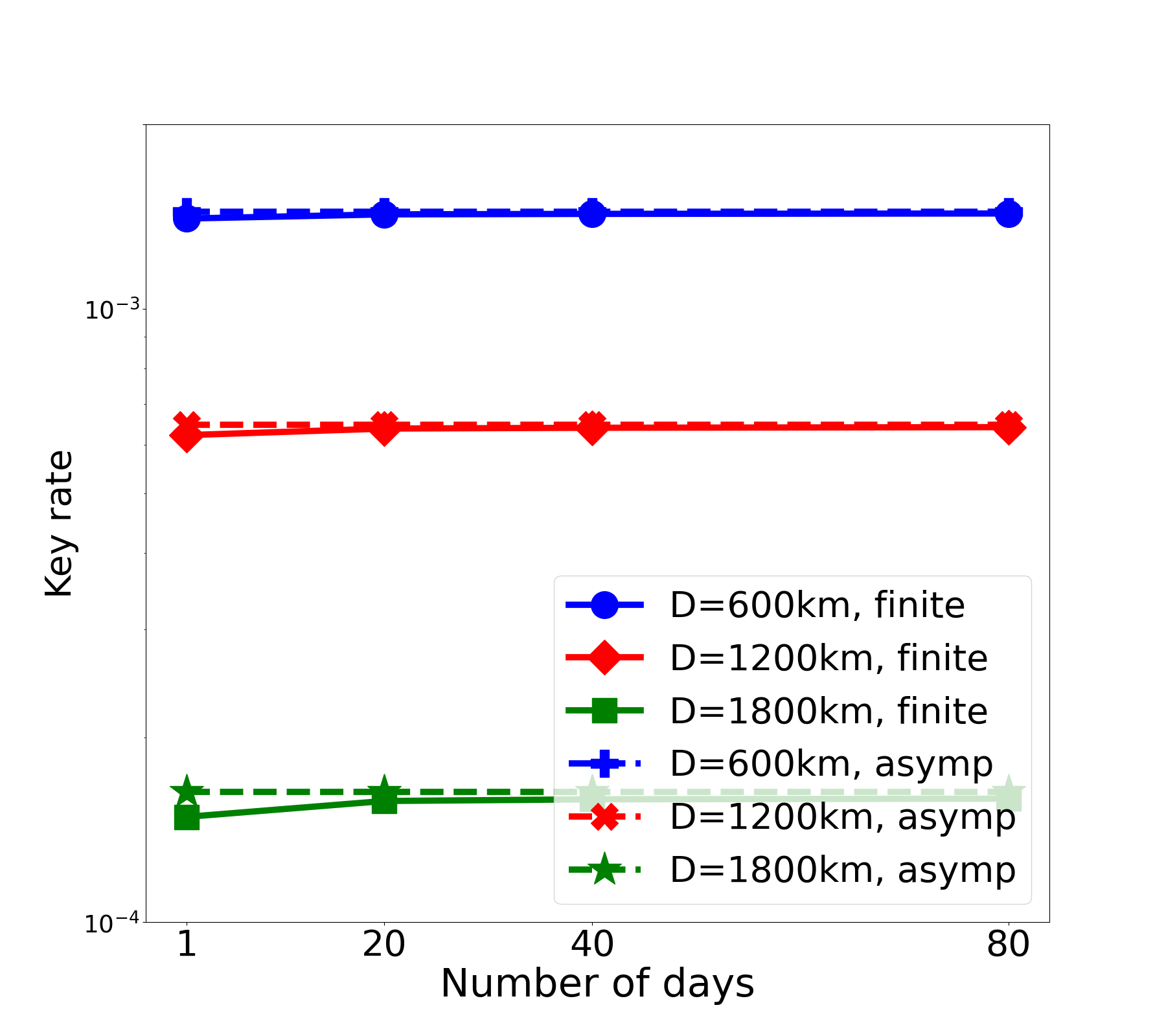}}
    \hspace{-.5cm}
    \subfigure[$A=800$ km]{\includegraphics[width=0.33\textwidth]{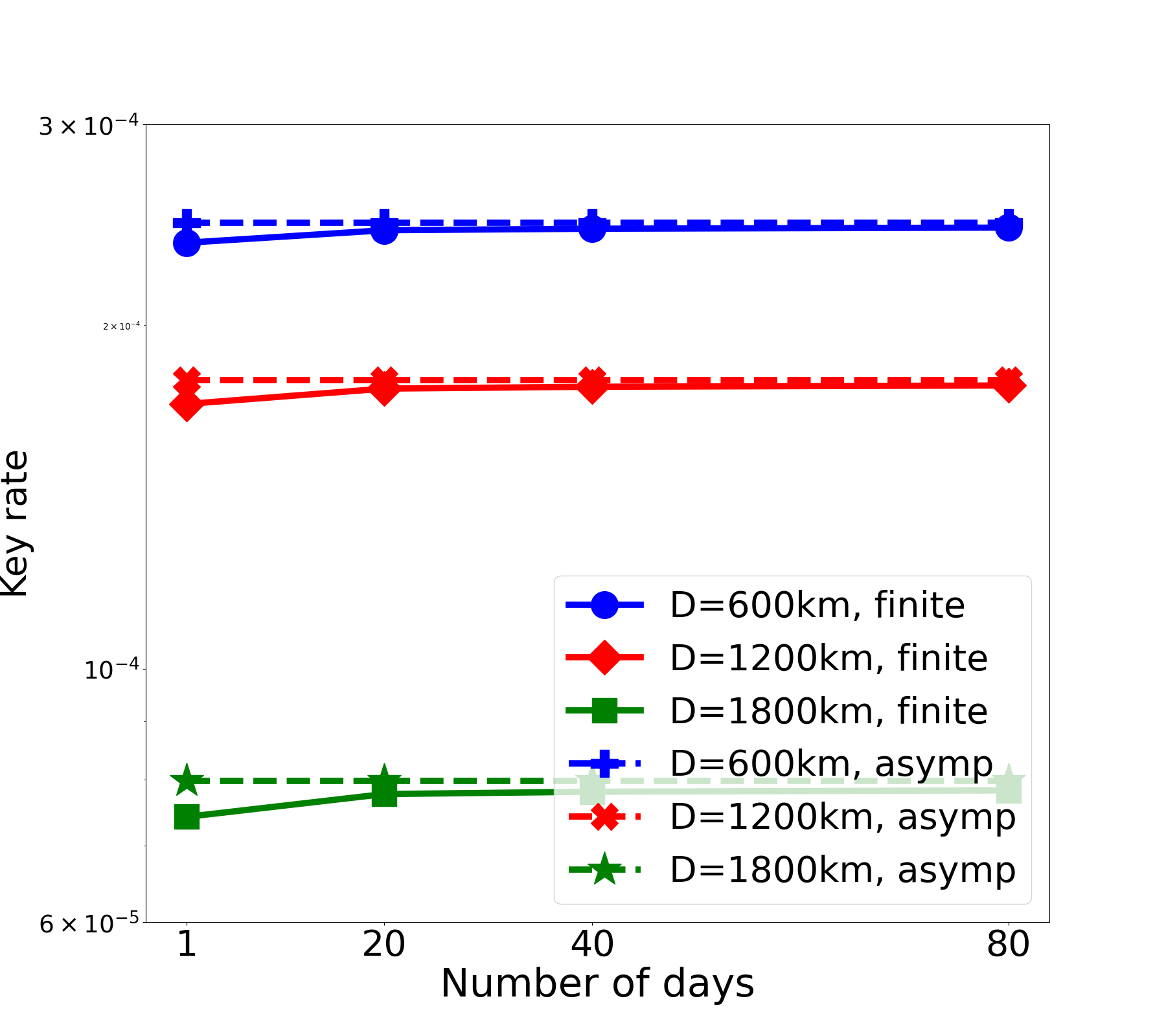}} 
    \hspace{-.5cm}
    \subfigure[$A=1000$ km]{\includegraphics[width=0.33\textwidth]{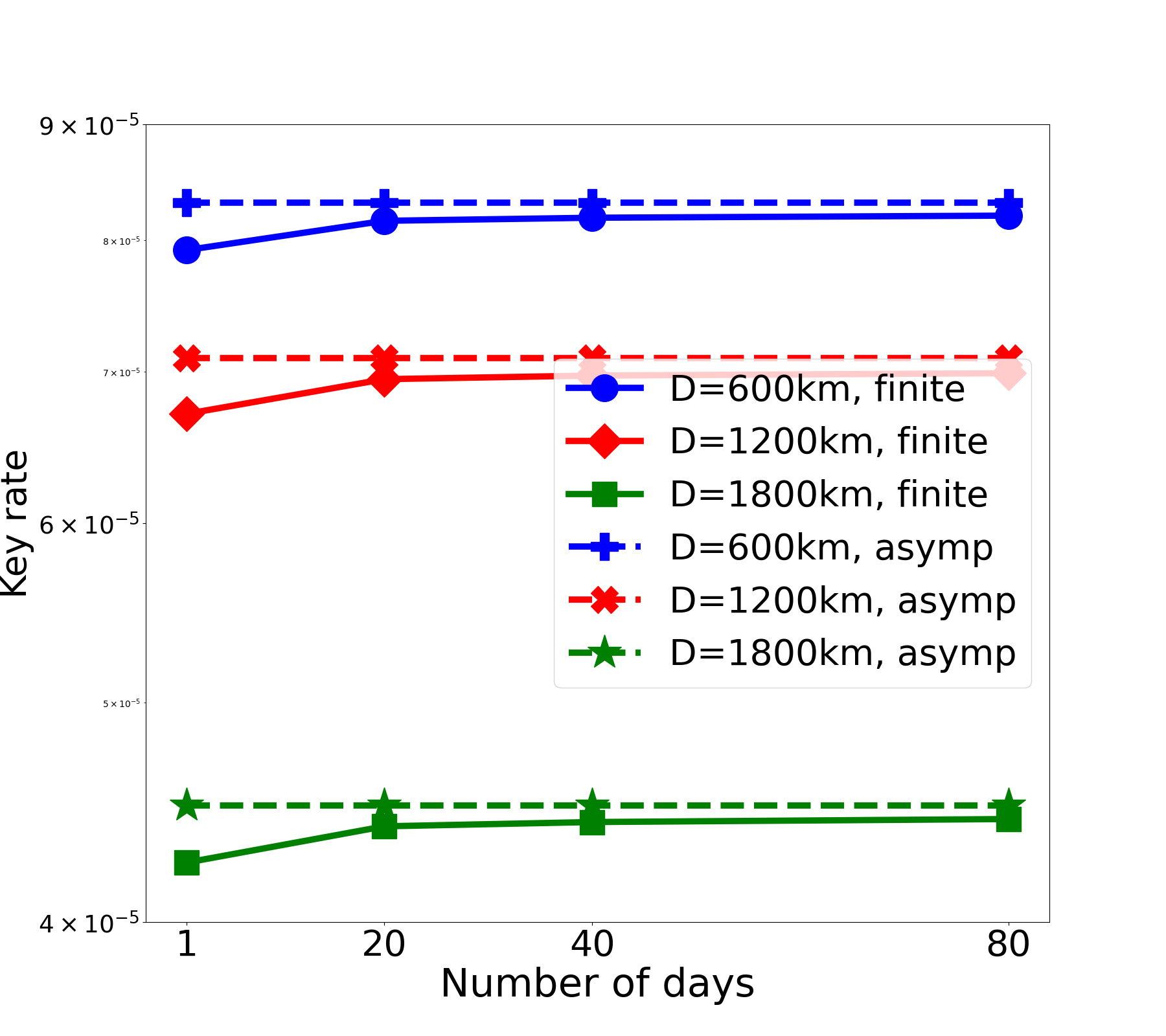}}
    \hspace{-.5cm}
    \caption{Key rate of the blockwise scheme for both finite and asymptotic scenarios. }
    \label{fig:eff_key_rate_alt}
\end{figure*}

\begin{figure*}[ht]
    \centering
    \subfigure[$A=500$ km]{\includegraphics[width=0.33\textwidth]{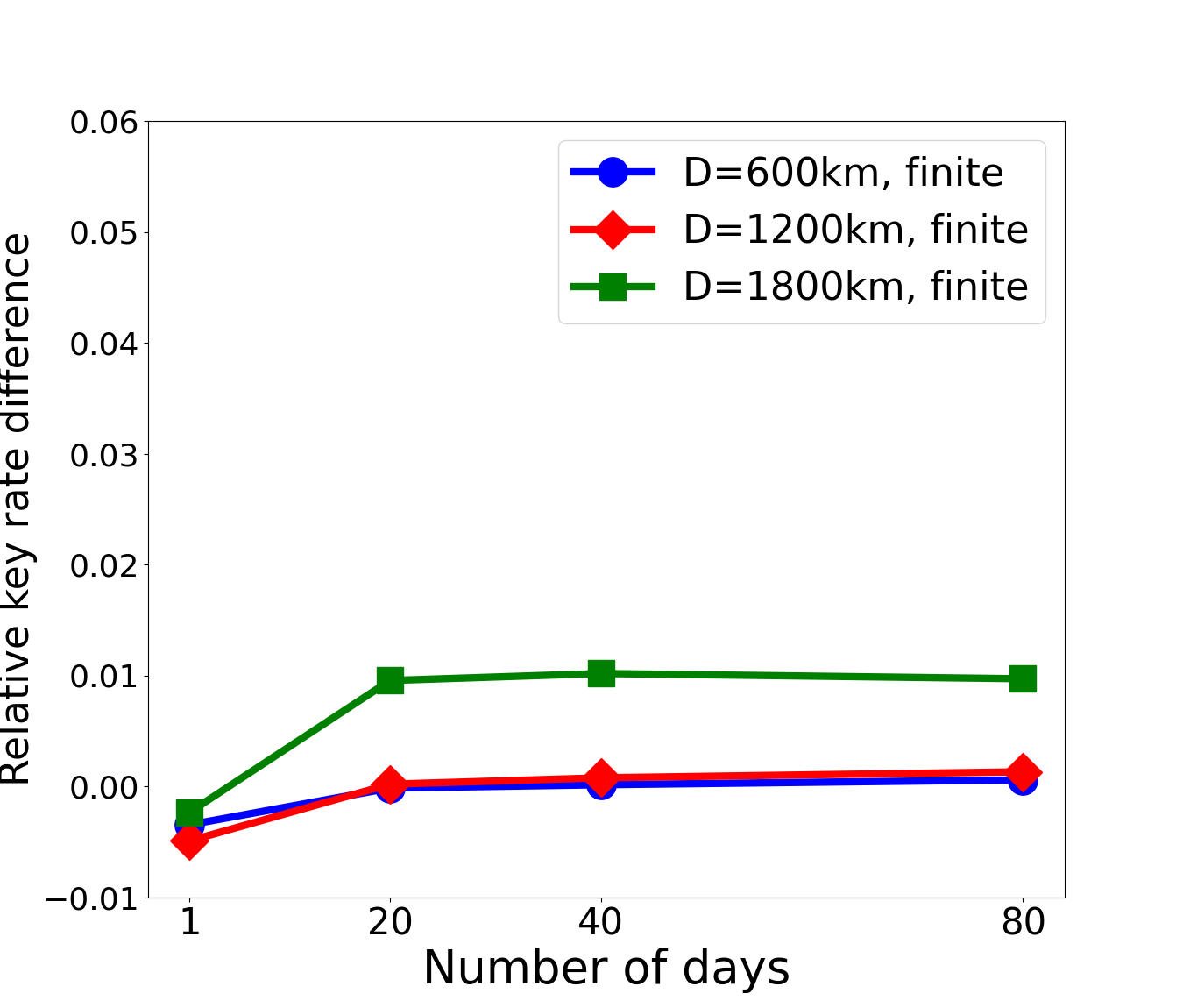}}
    \hspace{-.5cm}
    \subfigure[$A=800$ km]{\includegraphics[width=0.33\textwidth]{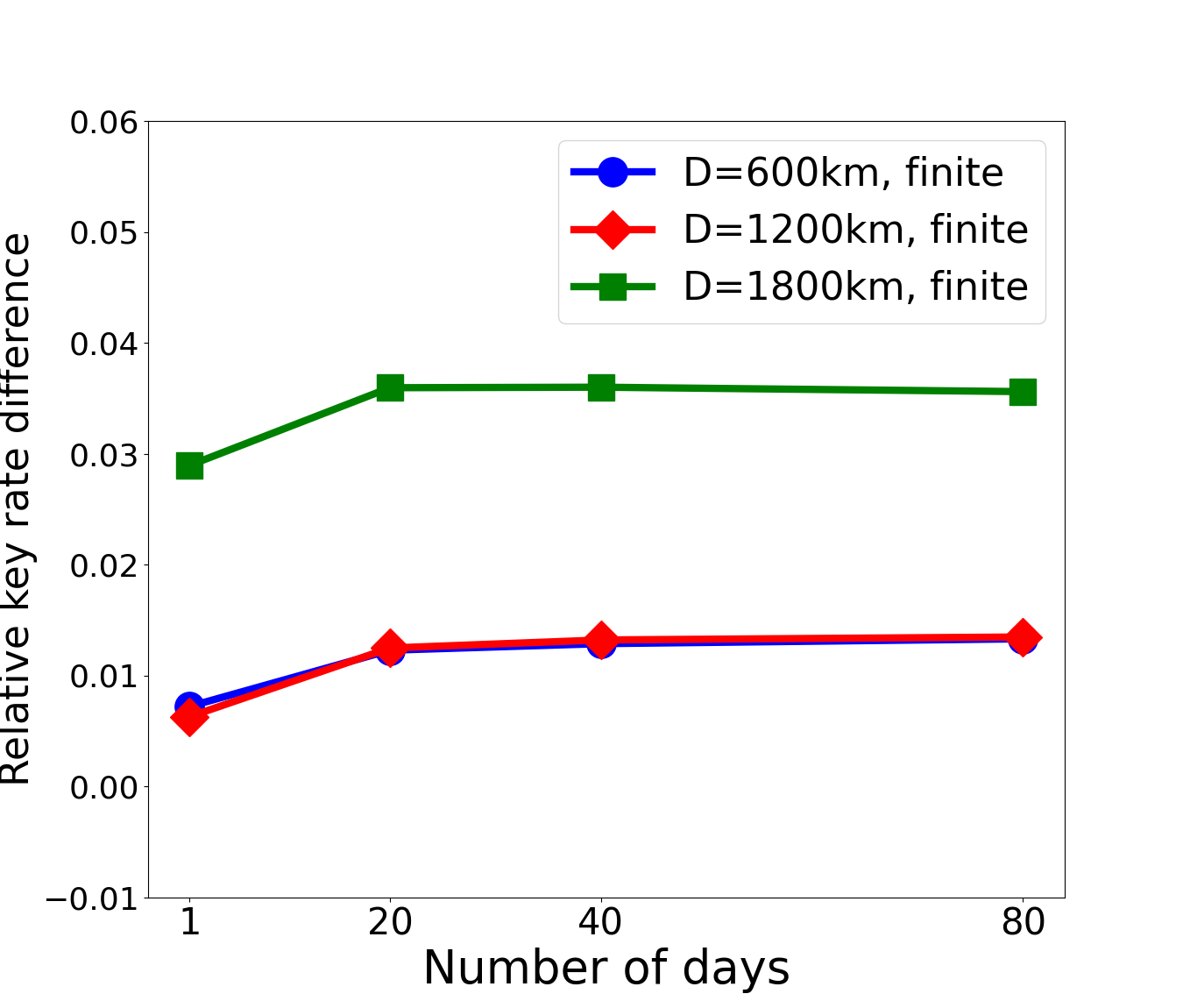}} 
    \hspace{-.5cm}
    \subfigure[$A=1000$ km]{\includegraphics[width=0.33\textwidth]{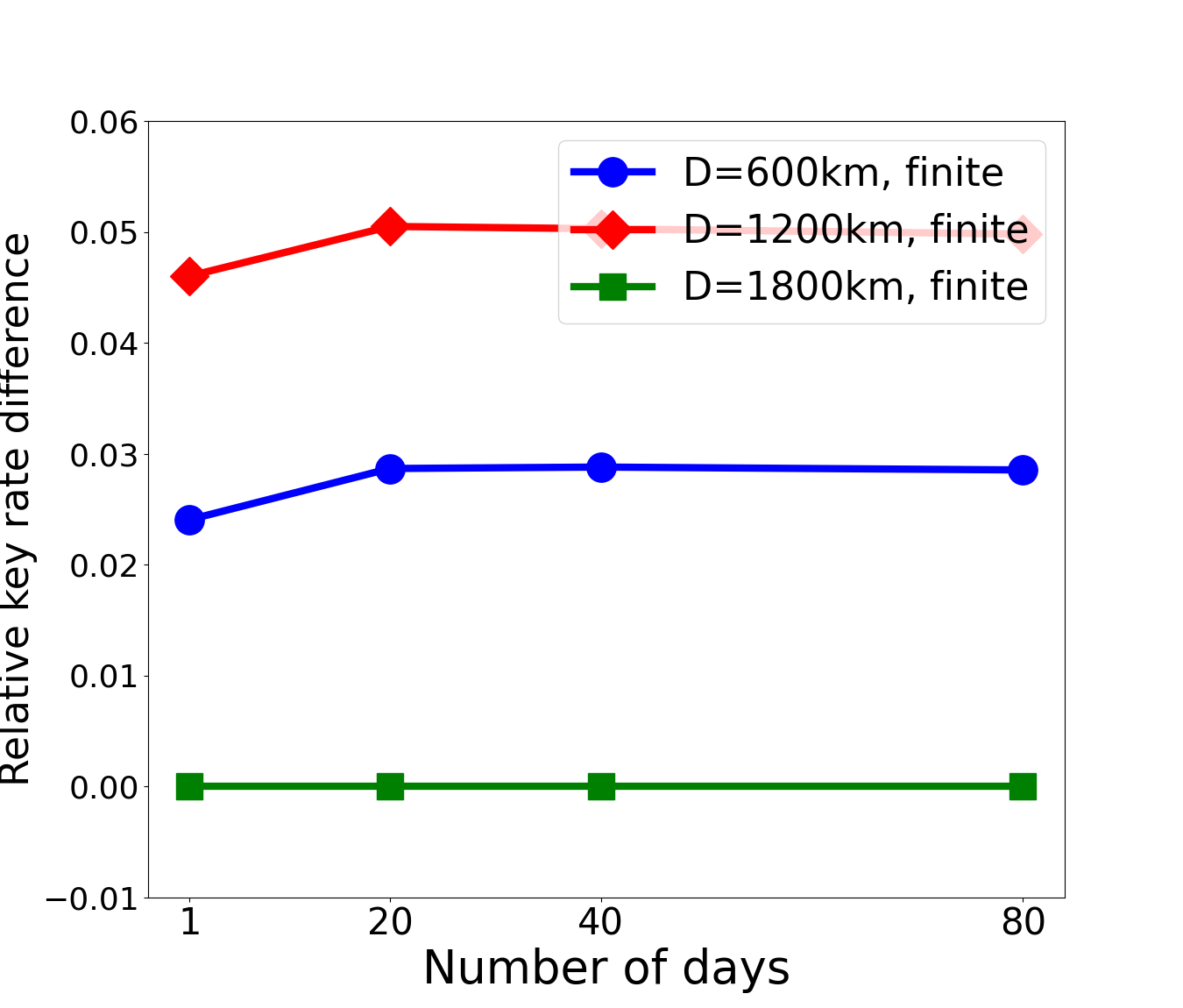}}
    \hspace{-.5cm}
    \caption{Relative key rate difference  between the blockwise and non-blockwise schemes. }
    \label{fig:relative_eff_key_rate_diff_alt}
\end{figure*}

We now compare the key rate of the blockwise
and non-blockwise schemes. Specifically, we assume that the schemes run over $k$ days, and set $k$ to 1, 20, 40, 60, and 80. 
For each $k$, we again select  the pump power and the sampling rate to maximize the number of secret keys generated over $k$ days. We see that the optimal pump power for $k$ days is similar to that of one day for both the blockwise and non-blockwise schemes (figure omitted).    

Figures~\ref{fig:eff_key_rate_alt}(a)-(c) plot the key rate under the blockwise scheme
when the satellite altitude is 500, 800, and 1000 km, respectively. In each plot, we show results for both the finite and asymptotic scenarios. We see that for the key rate for the finite scenario increases with the number of days, and approaches the asymptotic result when $k\ge 20$. 

We define $(r_b -r_{nb})/r_{nb}$ as the relative key rate difference between the blockwise and non-blockwise schemes, where $r_b$ and $r_{nb}$ are the key rate of the blockwise and non-blockwise schemes for the finite scenario, respectively. Fig.~\ref{fig:relative_eff_key_rate_diff_alt} plot the 
relative key rate difference for the various settings. 
We see that the difference is larger than 0 for all the cases when the satellite altitude is $800$ and $1000$ km, i.e., the blockwise outperforms the non-blockwise scheme. Specifically, the blockwise scheme leads up to 4\% and 5\% improvements when the satellite altitude is 800 km and 1000 km, respectively. When the satellite altitude is 500 km, we see up to 1\% difference between these two schemes, and the blockwise scheme leads to slightly lower key rate in some settings ($D=600$ and 1200 km and when the number of days is small). We only see four cases where the blockwise strategy leads to fewer key bits than the non-blockwise strategy: $A=500$km and $D=600$, 1200 or 1800 km, $k=1$ day; and when $A=500$ km, $D=1200$ km, and $k=20$ days. 

Let $\bar{\ell}_{block}$ and $\bar{\ell}_{non-block}$ represent the average number of key bits generated per day for the blockwise and non-blockwise strategies, respectively. Table~\ref{table:diff-number-of-key} shows $\bar{\ell}_{block}-\bar{\ell}_{non-block}$, where both quantities are obtained from the results of 80 days (i.e., the number of secret keys generated over 80 days divided by 80). We see that the blockwise strategy leads to $10^6$ to $1.9\times 10^7$  more keys per day in the various settings, except for one setting ($A=1000$ km and $D=1800$ km) since no key is generated during daytime for both strategies. 

Summarizing the the results in Fig.~\ref{fig:relative_eff_key_rate_diff_alt} and Table~\ref{table:diff-number-of-key}, we see that the blockwise strategy in general leads to higher key rate and more key bits except for the scenarios with low satellite altitudes and small number of days.
Therefore, it is in general more advantageous to use the blockwise strategy, which can be easily deployed since it is only used in the classical post-processing stage of QKD.

\begin{table}[t]
\centering
\caption{Average number of key bits generated using the blockwise scheme per day minus that generated using the non-blockwise scheme.}
\begin{tabular}{ |c||c|c|c|  }
 \hline
 & $D=600$ km & $D=1200$ km & $D=1800$ km\\
 \hline \hline
$A=500$ km & $2.9\times 10^{6}$  & $1.7\times10^{6}$ &  $10^{6}$ \\
 \hline
$A=800$ km & $1.8\times 10^{7}$  & $10^{7}$  &   $7\times 10^{6}$ \\
\hline
$A=1000$ km  & $1.6\times 10^{7}$     & $1.9\times 10^{7}$ &   $0$ \\
\hline
\end{tabular}
\label{table:diff-number-of-key}
\end{table}

\subsection{Handling Spurious 2-photon Terms} \label{sec:multi-photon}

 Recall that $p(2)$ in Eq.~(\ref{eq:spdc_state}) is the probability of generating a 2-photon term in
each pair of mode. Such 2-photon events are detrimental to QKD due to photon-number-splitting (PNS) attacks~\cite{Brassard00:PNS}. Specifically, when two photons (instead of one photon in an entanglement pair) are sent from the satellite to a ground station, an adversary can keep one photon and sends the other to the ground station, and hence knows the state at the ground station. So far, 
we have ignored $p(2)$ for ease of analysis. To investigate the impact of this approximation on our results, we 
simulate a hypothetical idealistic entanglement source, where either vaccuum  state or entanglement pairs are generated, i.e., we normalize $p(0)$ and $p(1)$ as $p(0)/(1-p(2))$ and $p(1)/(1-p(2))$, respectively, and then set $p(2)=0$. 
After that, the simulation of loss and noise on the entanglement pairs follows the models in Section~\ref{sec:prelim}. 
Then for the optimal pump power chosen in Section~\ref{sec:opt-pump-power}, we compare the resultant success probability and fidelity of this idealistic source with those of the actual SPDC source we use. We observe that these two sources have similar success probabilities in all settings (the difference is within 0.001). For fidelity, although their differences are small (within 0.01) in most cases, the difference can be large (0.03) when the satellite altitude is 500 km and the ground station distance is 600 km. Further exploration on such cases is left as future work.

\section{Related Work}

Satellite-based quantum communication provides a promising direction for global-scale QKD~\cite{rarity2002ground,peng2005experimental,ursin2007entanglement
, yin2012quantum
, ma2012quantum
, wang2013direct
, nauerth2013air
, yin2013experimental
, vallone2015experimental,liao2017satellite}. A recent study~\cite{Sidhu22:finite-key-sate} explores the finite key effect in satellite-based QKD. It considers a  satellite communicating with a single ground station, instead of entanglement-based QKD where a satellite transmits entanglement pairs to a pair of ground station simultaneously as in this study. In addition, it concatenates all the data together, i.e., it only considers the non-blockwise strategy, while our study compares blockwise and non-blockwise strategies. The authors use the finite key analysis techniques proposed in~\cite{Lim14:Concise,Yin20:Tight,Tomamichel17:Fundamental}. We derive our finite key results based on~\cite{tomamichel2012tight}. In particular, that reference provides tight key-rate bounds, using entropic uncertainty, when processing a raw key into a secret key.  Typically this method is used directly for the non-blockwise scenario which is usually considered in QKD research.  We also use their methods in our work to analyze the amount of secret key material in smaller blocks, running privacy amplification independently on each block and, thus, using results in~\cite{tomamichel2012tight} to determine the size of the secret key derived from each (smaller) block.  It would be interesting future work to see if one could bound the quantum min entropy of each sub-block directly and run a single privacy amplification processes over the entire block. That is, use a single invocation of privacy amplification, as in the non-blockwise strategy, yet still retain the benefit of increased key lengths as in blockwise postprocessing.
The loss and noise models in this paper are based on those in~\cite{panigrahy2022optimal}, and we extend its noise model by considering unfiltered background photons. The focus of~\cite{panigrahy2022optimal} is on optimal scheduling of satellite to ground station transmissions 
with a constellation of satellites. This work focuses on comparing blockwise and non-blockwise key distillation in satellite-based QKD.

\section{Conclusion and Future Work}
In this paper, we 
compare blockwise and non-blockwise key distillation strategies for satellite-based QKD, where the satellite quantum channel is highly dynamic and hence can produce raw key blocks with significantly difference characteristics. Using extensive simulation, we show that the blockwise strategy can lead to a 5\% higher secret key rate 
than the traditional non-blockwise strategy that is agnostic to the dynamics of the quantum channel.   

As future work, we will consider scenarios with multiple satellites in a constellation and multiple ground station pairs. We will also consider more factors when modeling quantum satellite channels (e.g., weather conditions, cloud coverage). In addition, we will consider more blocks based on time of the day (e.g., sunset, night, sunrise, noon).



\section*{acknowledgments}
\noindent This research was supported in part by the NSF grant CNS-1955744, NSF-ERC Center for Quantum Networks grant EEC-1941583,  MURI ARO Grant
W911NF2110325, and NSF CCF-2143644.

\balance
\bibliographystyle{ieeetr}
\bibliography{references,bib/quantum}
\end{document}